%% file: template.tex
\documentclass{article}

\usepackage{arxiv}
\usepackage{appendix}
\usepackage{chngcntr}
\usepackage{hyperref}  
\usepackage{xcolor}
\hypersetup{backref=true,       
    pagebackref=true,               
    hyperindex=true,                
    colorlinks=true,                
    breaklinks=true,                
    urlcolor= black,                
    linkcolor= blue,                
    bookmarks=true,                 
    bookmarksopen=false,
    filecolor=black,
    citecolor=blue,
    linkbordercolor=blue
}
\usepackage[utf8]{inputenc} 
\usepackage[T1]{fontenc}    
\usepackage{url}            
\usepackage{booktabs}       
\usepackage{amsfonts}       
\usepackage{nicefrac}       
\usepackage{microtype}      
\usepackage{lipsum}		
\usepackage{graphicx}
\usepackage{natbib}
\usepackage{doi}
\usepackage{appendix}
\usepackage{gensymb}
\usepackage{subcaption}
\usepackage{paralist}

\usepackage{amsmath}
\usepackage{makecell}

\input{notation}

\newcommand{\G}{\mathcal{G}}

\title{Learning earthquake sources using symmetric autoencoders}

\date{\today} 					

\author{ \href{http://orcid.org/0000-0003-4081-8969}{\includegraphics[scale=0.06]{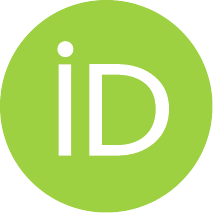}\hspace{1mm}Pawan Bharadwaj}\\
	Centre for Earth Sciences\\
	Indian Institute of Science\\
	Bengaluru \\
	\texttt{pawan@iisc.ac.in} \\
        \And
	Madhusudan Sharma \\
	Centre for Earth Sciences\\
	Indian Institute of Science\\
	Bengaluru \\
	\texttt{md24sharma@gmail.com} \\
	\AND
	\hspace{1mm}Isha Lohan \\
	Centre for Earth Sciences\\
	Indian Institute of Science\\
	Bengaluru \\
	\texttt{lohanisha@gmail.com} \\
	\And
        Pragna Sahoo\\
    Centre for Earth Sciences\\
	Indian Institute of Science\\
	Bengaluru \\
	\texttt{pragnap@iisc.ac.in} \\
}

\renewcommand{\shorttitle}{Learning earthquake sources using symmetric autoencoders}
\newcommand{\s}{\mathbf{s}} 
\newcommand{\coh}{\mathbf{s}} 
\newcommand{\p}{\mathbf{p}} 
\newcommand{\nui}{\mathbf{p}} 
\newcommand{\m}{\mathbf{m}} 
\newcommand{\KL}[2]{D_{\mathrm{KL}}\left(#1 \parallel #2\right)} 
\newcommand{\KLk}{H} 

\newcommand{\g}{\mathbf{g}} 
\newcommand{\f}{\mathbf{f}} 

\hypersetup{
pdftitle={Learning earthquake sources using symmetric autoencoders},
pdfsubject={q-bio.NC, q-bio.QM},
pdfauthor={Pawan Bharadwaj},
pdfkeywords={earthquake, time reversal, autoencoder, scattering, dereverberation, symmetry, rupture},
}

\begin{document}
\maketitle

\begin{abstract}
This study examines almost thirty deep-focus earthquakes, 
magnitudes starting from Mw 6.0 and higher,
with the aim of accurately determining the source-time function (STF) of P arrival and its azimuthal dependence.
We use the variational symmetric autoencoder (SymVAE), a neural network architecture designed to automatically isolate earthquake information from far-field seismic waves.
Our findings demonstrate that the STFs produced by the network uncover weak secondary episodes in numerous earthquakes, providing evidence that the majority deep-focus earthquakes release bursts of seismic moment.
This groundbreaking study is the first to generate high-resolution STFs without requiring traditional
path-effect deconvolution, a process that usually introduces substantial uncertainties and hinders achieving high temporal resolution.
Our unsupervised learning method for obtaining STFs does not require labeled seismograms and is based on the principle of scale separation, which allows the
accumulation of earthquake information from nearby receivers.
This principle states that the
variations in far-field band-limited seismic measurements resulting from finite faulting occur across two spatial scales: a slower scale associated with the source processes and a faster scale corresponding to path effects.
This research compares the STFs obtained from SymVAE with those gathered by stacking envelopes and traditional deconvolution. We show that the principal features of SymVAE-derived STFs align with those of envelope stacking, especially in dense seismic networks that record large-magnitude earthquakes with high signal-to-noise ratios.
However, SymVAE can effectively extract source details from fewer stations or in situations with low signal-to-noise ratios, where envelope stacking may not work; this capability is vital for directivity analysis.
%
We evaluated the quality of SymVAE output and performed a synthetic experiment to recover the source in the presence of path scattering.

\end{abstract}

\section{Introduction}



\begin{figure}
    \centering
    \includegraphics[width=1\linewidth]{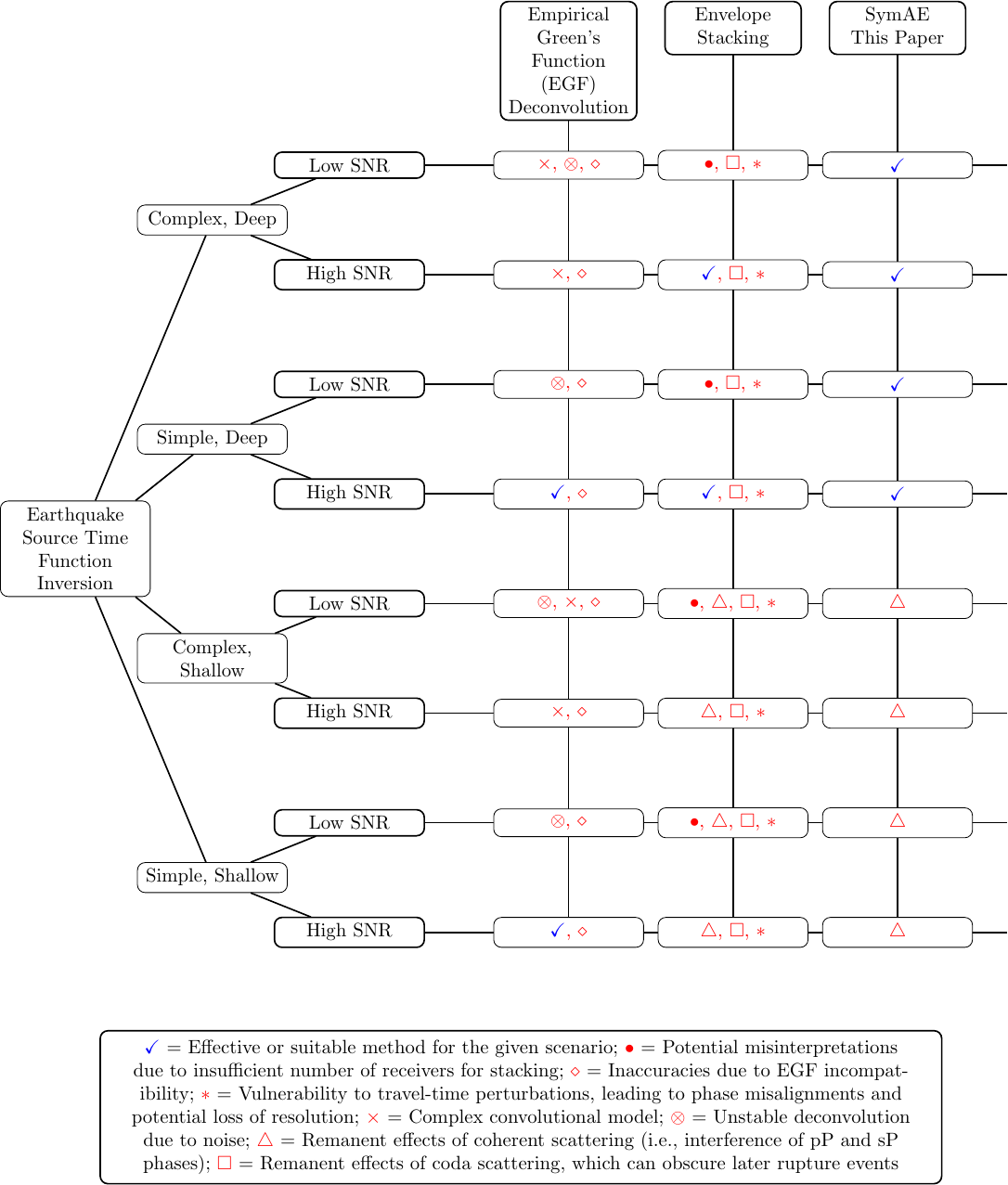}
    \caption{This figure illustrates the tree for earthquake source time function (STF) inversion. It categorizes scenarios based on the complexity (simple/complex) and depth (shallow/deep) of the earthquake, as well as the Signal-to-Noise Ratio (SNR). Horizontal lines represent different SNR conditions, while vertical lines indicate the methods used for inversion. Distinct markers at intersections highlight specific outcomes or issues related to each method.}
    \label{fig:decon_tree}
\end{figure}

Traditional far-field earthquake characterization comprises parameters whose inference is less affected by noise and scattering due to subsurface heterogeneity and is confined to longer periods.
For example, 
\begin{inparaenum}
\item the polarity of the first arrival is used to infer the focal mechanism i.e., mean seismic moment density tensor~\citep{amdzi_gcmt1981,EKSTROM20121};
\item the low-frequency level of the displacement spectra is used to estimate the moment magnitude~\citep{kanamori_2004_physics};
\item the approximate duration of P-wave arrivals to perform directivity analysis and infer rupture velocity~\citep[]{park2015inversion};
\item mean source spectra estimation by averaging over all the available stations \citep[]{prieto2004}.
\end{inparaenum}
Enhancing our understanding of earthquakes relies on techniques that accurately and with a high resolution extract angle-dependent apparent source time function (STF).
 Accurate source information is particularly important when estimating the kinematic rupture properties and directivity effects, described in \cite{madariaga_seismic_2015}.
However, 
inverting the STF from the observed seismic data can be challenging due to various factors such as the complexity of the source, subsurface scattering, and noise levels.
Here are brief explanations for each of the factors mentioned:
\begin{enumerate}
\item Complexity of the convolutional model. In some situations, e.g. in the case of multiple rupture episodes, heterogeneous fault structures, or variable rupture dynamics,
the assumption that the observed seismic signal is a straightforward single-path convolution of the source may not hold.
\item Additive noise and deconvolution instability. When noise is present in seismic data, attempting to perform a deconvolution to separate the source from the path effects can become unstable. Noise can amplify during the deconvolution process, leading to unreliable results.
\item Effects of coda scattering on weak secondary episodes. This scattering can obscure the signal needed to accurately identify stopping phases in STFs. For example, if one rupture episode's coda overlaps with another, it becomes challenging to distinguish between them.
\end{enumerate}
In Fig.~\ref{fig:decon_tree}, we present a comprehensive analysis of the methods used to extract earthquake STFs, where we categorize the inversion scenarios based on the simplicity or complexity of the source and the depth of the earthquake, as well as the signal-to-noise ratio (SNR) of the observed data. 
The figure begins by classifying the earthquake source as either simple, with a single rupture episode, or complex, with multiple rupture episodes. In addition, it classifies them as shallow or deep. For shallow earthquakes, the time difference between the P arrival and the associated surface-reflected arrivals is small, and we cannot window individual arrivals. However, for deep earthquakes, the time difference is large, allowing us to window P arrivals.
Each category is further divided on the basis of the SNR into high and low SNR scenarios. For each combination, we explore three different inversion techniques: empirical Green's function (EGF) deconvolution, envelope stacking, and SymVAE (symmetric autoencoder), the latter being the focus of this paper.




%

%

%

The empirical Green function (EGF) deconvolution technique, introduced by \cite{hartzell1978earthquake}, is particularly effective in situations characterized by high SNR, simple, and deep earthquake sources~\citep{irikura1986,plourde_multichannel_2017-1,wu_source_2019,lanza_source_1999,hutchings2012application,bezada2012,zhan_supershear_nodate}.
EGFs are seismograms that have a nearly impulsive source function and path effects similar to the main earthquake. 
The majority of techniques derive empirical Green functions (EGFs) from low-magnitude earthquakes originating from an identical source region. This approach offers the advantage that a smaller earthquake might share similar coda scattering effects with the main earthquake, allowing for deconvolution. However, several limitations are associated with this method. Consider the following scenarios: \begin{inparaenum} \item if the EGF seismogram has a low signal-to-noise ratio (SNR) because of small magnitude, deconvolution becomes inefficient~\citep[]{10.1111/j.1365-246X.1976.tb01267.x}. On the other hand, picking an intermediate magnitude earthquake with a better SNR might lead to EGF incompatibilty~\citep{Alessio1990}; \item when the main earthquake has few aftershocks, this method cannot be applied. \item 
if the chosen EGF is a complex earthquake rather than a point source, it will emit varying apparent source functions across different azimuths, potentially complicating the directivity analysis of the primary earthquake.
\end{inparaenum}
Some methods employ synthetic simulations with established Earth models to perform deconvolution, as demonstrated by \cite{vallee_scardec_2011, vallee_new_2016-1}. This approach avoids the issue of EGFs possessing a low SNR; however, generating the Green function in the scattering regime poses difficulties, mainly because the Earth models used are limited to low wavenumbers.
Another major issue with these deconvolution methods is that they are not array-based, that is, they operate individually on seismograms.

One significant drawback of the EGF method is its unsuitability for examining shallow earthquake sources due to the invalidity of a simple one-input, one-output convolutional model~\citep{langston1975procedure, andrews1986objective}
\begin{eqnarray}
d(\x, \omega)  \approx G(\x, \omega)\,s(\x, \omega),
\end{eqnarray}
where the 
measured displacement $d$ at spatial location $\x$ is approximately given by a multiplication in the frequency domain (convolution in time) between an apparent source function $s$ and a path-dependent Green's function $G$.
We used $\omega$ to denote the temporal frequency.
%
%
This simple convolutional model ignores the sensitivity of the apparent source function to the type of seismic arrival, e.g., P, pP, PP, or sP ~\citep{bormann2013iaspei}.  Contrarily, in fact a seismogram can be more accurately modeled after decomposing Green's function into, e.g., $G \approx G_{\text{P}} + G_{\text{pP}} + G_{\text{sP}} + \cdots$, using the high-frequency approximation to write 
\begin{eqnarray}
d(\x, \omega)  \approx G_{\text{P}}(\x, \omega)\,s_{\text{P}}(\x, \omega) + G_{\text{pP}}(\x, \omega)\,s_{\text{pP}}(\x, \omega) + G_{\text{sP}}(\x, \omega)\,s_{\text{sP}}(\x, \omega) + \cdots.
\label{eqn:gseries1}
\end{eqnarray}
This leads to a convolution+mixer model, where several apparent source functions ($s_{\text{P}}$, $s_{\text{pP}}$ and $s_{\text{sP}}$) are mixed after undergoing unique path-specific convolutions. 
Likewise, the EGF method is ineffective for handling complex earthquake sources with multiple ruptures on various fault planes, as
the single-path convolution model assumes that the seismic moment density is uniform (up to a scalar multiple) in the source region. A more accurate representation of a complex earthquake involves mixing rupture episodes with different source functions, denoted as $s^1$, $s^2$, $s^3$, and so on:
\begin{eqnarray}
d(\x, \omega)  \approx G^{1}(\x, \omega)\,s^1(\x, \omega) + G^2(\x, \omega)\,s^2(\x, \omega) + G^3(\x, \omega)\,s^3(\x, \omega) + \cdots.
\label{eqn:gseries2}
\end{eqnarray}
Here, the convolution+mixer model acknowledges that the apparent source functions are convolved with distinct Green's functions ($G^1$, $G^2$, $G^3$, etc.) specific to the moment tensor of each rupture episode.
The array-based method introduced in this paper is similar to backprojection methods (e.g. \citet{Ishii2005, Kiser2012, zeng2022travel}), in the sense that it utilizes the spatial coherence of the wavefield recorded by a seismic network to analyze and extract the STFs of earthquakes. Our approach uses novel deep learning techniques to enhance extraction accuracy without assuming a single-path convolutional model.
Our framework generalizes multichannel blind deconvolution, substituting neural networks for the convolution signal model, thereby extracting STFs without the requirement for EGFs.
It uses a neural network architecture, termed a variational symmetric autoencoder~\citep[SymVAE]{bharadwaj2024extractingcoherentseismicwavefield}, capable of being trained to extract coherent information from far-field seismograms.
We demonstrate that SymVAE can be effectively applied to complex earthquake sources involving multiple rupture events, as well as to low-SNR situations with earthquakes having a magnitude around 6.0 Mw.
Cepstral domain techniques serve as an example of methods for the blind estimation of source time functions; nevertheless, to reconstruct phase spectra, these techniques frequently presume the presence of minimum phase source pulses~\citep{ulrych_homomorphic_1972}.
%

%
%

%

%

%

Envelope stacking does not require a single-path convolutional model, making it suitable for more complex earthquakes compared to the EGF method. For directivity analysis~\citep[]{https://doi.org/10.1002/2015GL064587}, this method stacks several seismograms from the same azimuth using their envelope functions, to reduce coda scattering and lower additive noise.
Generally, the availability of teleseismic receivers is limited for the majority of the azimuths, making this approach primarily effective in high-SNR situations.
Moreover, fluctuations in the quantity of seismograms and the quality of their signals as a function of azimuth can result in variations in the envelope stacks, potentially resulting in misinterpretations and incorrect associations with directivity effects.
A significant downside of envelope stacking is that it is
vulnerable to travel-time perturbations, as it relies on the precise initial alignment of seismograms to enhance coherent source signatures ---
in this paper, we generated stacked envelopes by aligning displacement seismograms using the IASP91 model. 
Travel-time perturbations exceeding a quarter of the period can introduce phase misalignments, leading to signal interference during stacking and potentially causing a loss of resolution. 
This approach cannot be utilized for shallow earthquakes, even in cases of high SNR, as it does not involve deconvolution. Consequently, the influence of coherently scattered surface-reflected arrivals such as pP and sP remains present in envelope stacks.
SymVAE resembles envelope stacking in that it also avoids deconvolution and is ineffective in eliminating the impact of coherently scattered arrivals.
 However, SymVAE can automatically correct for travel-time perturbations, is not affected by azimuthal variation of receiver distribution, and can isolate the influences of additive noise and coda scattering, thus yielding higher-resolution STFs even in low-SNR conditions for deep-focus earthquakes, regardless of their complexity.

%




\begin{figure}
     \centering
         \includegraphics[width=0.8\textwidth]{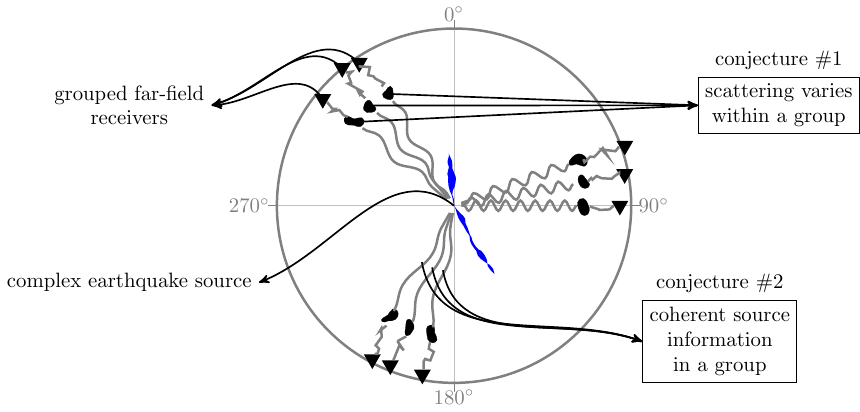}
                        \caption{
Symmetric autoencoder assumes a separation between the source and the path scales i.e., 
the earthquake source information 
(depicted using the raypath pattern) varies
on a much slower spatial scale than the path effects (depicted using the blob shape).
Here, for simplicity, only the azimuthal angle is considered as spatial coordinate, and ray bending is ignored.
Each triangular marker represents a receiver.
}\label{fig:intro_scale}
\end{figure}


%

%
Our approach uses an adaptation of variational autoencoders~\citep{kingma2013vae, rezende2014stochastic, burda2015iwae, higgins2017beta, doersch2016tutorial} to
isolate spatial coherence in the measured seismic displacement field. 
Traditional autoencoders cannot be trained to disentangle the source information from seismograms.
The design of our architecture relies on the fundamental assumption of scale separation, which is depicted in Fig.~\ref{fig:intro_scale}, associating the spatial variations \emph{slow} and \emph{fast} with the phenomena of the source and the path, respectively.
It is crucial to clarify that the terms slow and fast do not refer to time but rather to spatial characteristics. Specifically, slow denotes spatial variations with lower rapidity, while fast corresponds to spatial variations with higher rapidity.
Scale separation states that the variations in the far-field seismic measurements due to finite faulting occur across two different spatial scales --- a slower scale in which the source processes occur and another faster scale corresponding to the path effects.
Scale separation leads to an approximate symmetry: source information is invariant with respect to receiver permutations within a group of \emph{closely-spaced} receivers. We validate scale separation using the representation theorem and the Fraunhofer approximation for finite fault sources.

SymVAE utilizes the symmetry discussed above and offers a valuable middle ground between single-station
analysis and backprojection methods.
It
scales well with the data and benefits from improvements in the spatial coverage of seismic networks. 
The training of SymVAE is unsupervised; therefore, our approach does not require travel time or source-duration picking, empirical Green's function, knowledge on the subsurface scattering, modeling of source physics and path effects, etc. 
Notably, SymVAE can be trained on all available far-field seismograms of multiple earthquakes.
In this study, displacement seismograms from approximately 30 complex deep-focus earthquakes~\citep{heidi2015} were used to extract source-time functions. These functions enabled the identification of weaker seismic moment release bursts, which can potentially support the hypothesis of multiple mechanisms~\citep{Chen1996,zhan2017gutenberg,zhan2020mechanisms}. This hypothesis suggests that various physical mechanisms might play a role in the progression of deep-focus earthquakes.
We conducted several validation tests to confirm that SymVAE successfully disentangles source and path effects and possesses the capability to generate meaningful
STFs.
The remainder of the paper is organized as follows: we begin by introducing the concept of scale separation in the next section, followed by a section outlining the preparation of the training dataset. We then delve into the architecture of symmetric autoencoders, and the final sections present the application of the trained network to both synthetic and real earthquake data.

\section{Scale Separation}
\label{sec:scale}


%
In this section, we use the Fraunhofer approximation to demonstrate that the spatial variation of the displacement field caused by finite faulting can be divided into two scales: a slow scale related to the source effects and a fast scale associated with subsurface scattering.
This process, called scale separation, enables us to interpret spatial coherence in the displacement field as a characteristic of the seismic source.
This demonstration of scale separation through the Fraunhofer approximation is preliminary; numerical modeling of complex rupture scenarios could be employed to study its validity, a task we defer to future work.
Our starting point is the representation theorem~\citep[Chapter~3]{aki2002quantitative}. 
We consider a 
displacement discontinuity at $\xs=(\xi_1,\xi_2,\xi_3)$ in the fault region.
The $i$th component of the far-field displacement $u^i$ thus generated at $\x=(x_1, x_2, x_3)$ is given in the frequency $\omega$ domain as:
\begin{eqnarray}
  \label{eq:rep}
	u^i(\x,\omega;\xs)=\sum_{j,k=1}^3
	\G^{ij,k}(\x,\omega;\xs)\,m^{jk}(\omega;\xs).
\end{eqnarray}
Here, 
$m^{jk}$ denotes the $(j,k)$th component of the moment density tensor; and $\G^{ij,k}$ denotes 
the $k$th spatial derivative of the $(i,j)$th component of the elastodynamic Green's tensor.
We first decompose Green's tensor into components associated with individual ray paths utilizing the high-frequency approximation: $\G = \G_1 + \G_2 + \cdots$, where the subscript denotes the ray path index.
Then, we only consider the ray paths having wave velocity $\alpha$ in the source region, for instance, P, pP, PP rays.
Finally, we rewrite Green's term in the Eq.~\eqref{eq:rep} after employing Fraunhofer approximation as
\begin{eqnarray}
  \label{eq:fraun}
  \G^{ij,k}(\x, \omega;\, \xs) \approx \sum_{l}
  \G^{ij,k}_{l}(\x, \omega;\, \xs_0)
   \exp{[-\imath\omega(\xs \cdot \gamma_{l})/\alpha]}, 
\end{eqnarray}
where $\gamma_l$ is the unit vector
that points in the direction of the $l$th ray
leaving $\xs_0$ toward the receiver.
The focus denoted by $\xs_0$ is the origin.
Note that a homogeneous P-wave velocity is assumed in the source region. 
The Fraunhofer approximation simplifies the full-wave solution for the displacement field caused by finite faulting by assuming that the fault is far enough from the receiver. 
This approximation accounts only for the far-field phase correction, or travel-time difference, between the source region locations $\xs$ and $\xs_0$ and the receiver.
The validation of the Fraunhofer approximation, which is integral to the concept of scale separation, was assessed by ray tracing in Appendix \ref{app:frf}, although it is recognized as a somewhat weaker form of validation. 
Note that the validity of the Fraunhofer approximation imposes constraints on the highest possible frequency. 

%

%
We can express the total displacement at $\x$ by integrating $u^i$ over the fault plane $\varXi$ using the principle of superposition and rewriting eqs.~\eqref{eq:rep} and \eqref{eq:fraun}, as follows:
\begin{eqnarray}
  \label{eq:fraunsum}
	d^i(\x,\omega)=\int_{\varXi}\,u^i(\x,\omega;\xs)\,\text{d}\varXi
  =\sum_{l}\sum_{j,k=1}^3 \left(\G^{ij,k}_{l}(\x,\omega;\xs_0) 
  \underbrace{\int_{\varXi} m^{jk}(\omega;\xs)\,\exp{[-\imath\omega(\xs \cdot \boldsymbol\gamma_l)/c]}\,\text{d}\varXi}_{\text{apparent source function}\,s^{jk}_{l}(\omega; \gamma_l)}\right).
\end{eqnarray}
Here, $\text{d}\varXi$ represents an infinitesimal element at $\xs$ in the source region.
In the equation above, we have noted the apparent source time function that we seek to determine in this paper. Evidently, the STF is specific to an individual seismic arrival, due to its dependence on $\gamma_l$, and it is also specific to the fault plane corresponding to the rupture episode.
Assuming that the moment density tensor remains consistent across each fault plane, we can derive both convolution+mixer scenarios previously mentioned. 
\begin{inparaenum}
    \item Leaving out the existence of multiple fault planes and taking multipathing into account, we arrive at the model depicted in Eq.~\eqref{eqn:gseries1}, which separates the contributions from different arrivals.
    \item Alternatively, by focusing on a single seismic arrival while accounting for multiple fault planes, a model incorporating contributions from multiple rupture episodes across various fault planes can be formulated, as shown in Eq.~\eqref{eqn:gseries2}.
\end{inparaenum}
We once more emphasize that employing a simplified convolutional model along with deconvolution can potentially restrict the precision of the source information extracted from the seismic wavefield.
Our approach, on the other hand, does not impose restrictions on the source model; instead, we rely on the principle of scale separation described below.

%



\subsection{Source Scale}
The extent of the fault region determines the spatial scale at which the source information varies. 
For example, take into account the scenario of a point source located in the focus, represented by $m^{jk}(\omega;\,\xs)=M^{jk}(\omega)\delta(\xs)$, where $M^{jk}$ is a function of the angular frequency $\omega$ and $\delta$ is the Dirac delta function. This scenario produces a source term that is independent of the take-off angle $\gamma_l$ of the ray.
Generally, 
we observe that the form of the source integral (in Eq.~\eqref{eq:fraunsum}) is the spatial Fourier transform of the moment distribution in the source region \citep{laurentnotes}.
Then, we apply the convolution theorem to determine the source scale.
For convenience, we assume a line source along the $s\xi_1$ dimension and also use a single arrival for simplicity by removing the subscript $l$.
Rewriting the source integral using these simplifications leads to:
\begin{equation}
s^{jk}(\omega; k_1) = \int_{\xi_1} \Pi\left(\frac{2\xi_1}{L_1}\right)\,m^{jk}(\omega;\xi_1)\,\exp{[-\imath\,k_1\,\xi_1]}\,\text{d}\xi_1,
\end{equation}
where $L_1>0$ is the length of the fault, $\Pi$ denotes a unit rectangular function and 
$k_1=\omega \cos(\psi) /c$
is the transform variable (wavenumber)
dependent on the angle $\psi$ between the 
line source 
($\xi_1$ axis) and the 
ray takeoff direction
(unit vector $\gamma$).
Here, the multiplication by $\Pi$ constrains the spatial extent of the moment rate function $m^{jk}$. According to the convolution theorem, this multiplication is equivalent to a 
convolution in the wavenumber domain with a sinc function
\begin{equation}
\label{eq:sinc}
\mathcal{F}\left(\Pi\left(\frac{2\xi_1}{L_1}\right)\right) = \frac{L_1}{2\sqrt(2\pi)}\,\mathrm{sinc}\left(\frac{L_1k_1}{4}\right),
\end{equation}
where $\mathcal{F}$ denotes Fourier transform.
This convolution smears the source information along the spatial dimension 
of the measurements --- it leads to a reduction in the 
resolution proportional to the reciprocal of the fault length $L_1$ \citep{harris1978use}.
The resolution along $k_1$, which can be approximated by the width of the $\mathrm{sinc}$ function in Eq.~\eqref{eq:sinc}, is given by
\begin{equation}
{\Delta}k_1 = \frac{4 \pi}{L_1}.
\end{equation}
Finally, a linear approximation of 
$k_1=\omega\cos(\psi)/c$, as given by
\begin{equation}
{\text{d}}k_1 = \frac{ - {\text{d}}\psi\,\omega\,\sin\left( \psi \right)}{c},
\end{equation}
with perturbations $\text{d}k_1$ and $\text{d}\psi$, can be used to 
estimate the corresponding angular resolution
\begin{equation}
\Delta\psi = L_1^{-1}\frac{4\,\pi\,c}{\omega\,\sin(\psi)}.
\end{equation}
The angular resolution $\Delta\psi$ for a three-dimensional source region depends on the length of the dominant dimension $L_{\text{max}} = \max\left\{L_1, L_2, L_3\right\}$. 
For large-magnitude earthquakes, assuming large fault lengths, the source information varies at a faster spatial scale. 
For a given fault length, the source information in S waves (high $c$) originating in the source region exhibits variations on a faster scale than P waves (low $c$).
We analyze frequencies up to $0.1\,$Hz and use Tab.~\ref{tab:resP} to determine the expected angular resolutions for various fault lengths for both P and S waves.
For example, if the dominant length of the source region is $300\,$km, the expected angular resolution of the source information of the S wave is roughly $17\degree$.
Our next objective is to show that the path information varies on a faster scale compared to this source scale, as depicted in Fig.~\ref{fig:intro_scale}.

\begin{table}
      \centering
\input{tablep2.tex}
\caption{\label{tab:resP}
The angular resolution of the source information is dependent on the dominant length of the source region, and in our study, we used the Healpix spherical tessellation of the focal sphere to group the receivers. The approximate angular resolution of this tessellation was $14\degree$.}
\end{table}

\subsection{Path Scale}
\begin{figure}
    \centering
         \includegraphics[width=0.8\textwidth]{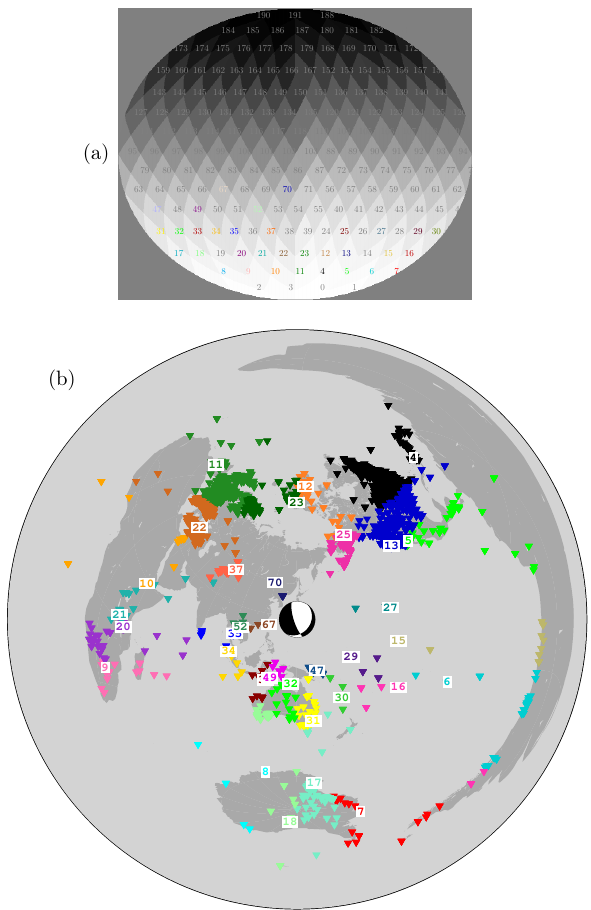}
        \caption{(a)
         Hierarchical Equal Area isoLatitude Pixelation \citep[HEALPix]{gorski2005healpix} of the earthquake focal sphere, where areas of all the pixels are identical, as the name suggests. The pixels are arranged in order of increasing azimuthal angle while maintaining constant polar angles, e.g., it can be observed that pixels 4--11 have the same polar angle.
         (b) Grouping of receivers: each numbered receiver group (represented by a distinct color) corresponds to a pixel on the focal sphere. The pixels on the focal sphere are also color-coded in the same unique colors to indicate their association with specific receiver groups.
         Here, we present an example of the Bonin earthquake (Mw 7.8) that occurred on May 30, 2015.
        \label{fig:pixels}
                 }
\end{figure}

\begin{figure}
\noindent\includegraphics[width=\textwidth]{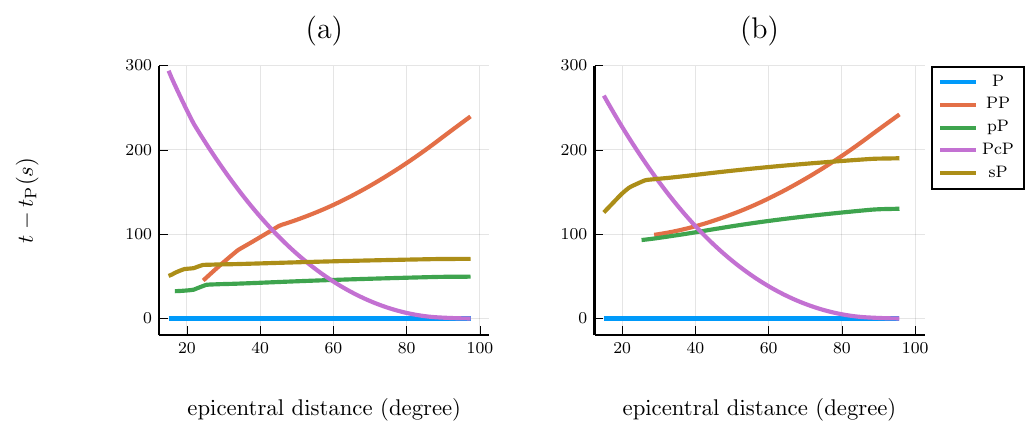}
\caption{
Traveltime delays relative to the P arrival for certain scattered arrivals in the P-window of seismograms are depicted for both (a) 200 km and (b) 600 km deep earthquakes.
Here, the arrivals such as PP and PcP are indicative of incoherent scattering, as their delays exhibit variations exceeding a quarter of the maximum period ($100\,$s), which corresponds to the low-frequency limit examined in our study.
Conversely, the surface-reflected arrivals pP and sP exhibit coherence, as their delays show minimal dependence on the epicentral distance.
Incoherent scattering aligns with the scale-separation conjecture (Fig.~\ref{fig:intro_scale}). On the other hand, surface-reflected coherent scattering poses a challenge as it cannot be disentangled from the source effects within the framework we have presented.
}
\label{fig:path_scale}
\end{figure}

Considering the minimal angular resolution of the source information as presented in Tab.~\ref{tab:resP}, we opted for the Hierarchical Equal Area isoLatitude Pixelation method \citep[HEALPIX]{gorski2005healpix}. HEALPIX divides the focal sphere from the earthquake into pixels, each with an equal surface area, providing an approximate angular resolution of $\Delta\psi=14\degree$. Fig.~\ref{fig:pixels}a illustrates the resulting pixelated focal sphere.
Within this pixelation scheme, all traced rays intersecting a particular focal-sphere pixel are assumed to carry identical source information. In other words, the source effects are considered coherent across all far-field receivers connected by rays intersecting a given pixel.

Within the context of focal-sphere pixelation, scale separation implies that the effects of coda scattering and multipathing notably vary among a set of receivers linked to a specific pixel.
An important observation is that the receivers linked to each pixel are distributed across a wide range of epicentral distances.
Moreover, the coloring of the receiver groups in 
Fig.~\ref{fig:pixels}b reveals that lower-indexed pixels exhibit receivers spread across a broader range of epicentral distances in comparison to higher-indexed pixels.
Considering the heterogeneity of the subsurface, it is plausible to presume that the effects of coda scattering differ substantially among the receivers within a specific pixel.

To establish scale separation in the case of multipathing, 
we selected a pixel and computed the raypaths connecting the source focus to the receivers while intersecting the pixel. Subsequently, we determined the corresponding travel times along these raypaths, and the results are presented in Fig.~\ref{fig:path_scale}.
Our analysis focused on examining the variation in the travel time delay of scattered arrivals relative to the P arrival within the receiver group.
Similar analysis can be performed for waves recorded in the S window.
This variation in traveltime delay among the receivers serves as an indicator of the scale at which multipathing effects fluctuate. Significant variations in time delay among the receivers suggest that path effects are changing at a faster rate than source effects, which remain uniform for the selected pixel.
Arrivals with a time-delay variation of less than a quarter of the period within the group of receivers are classified as coherently scattered arrivals. This category typically includes arrivals of pP and sP, as shown in Fig.~\ref{fig:path_scale}. It can be seen that the pP and sP arrivals arrive after a particular time delay following the P arrival irrespective of the epicentral distance, and their travel-time curves are roughly parallel to that of the P arrival.
On the other hand, arrivals with traveltime delays of more than a quarter of the period within the group of receivers are classified as incoherent scattering. This category includes crustal phases, PP, PcP in the case of multipathing, and coda scattering due to random inhomogeneities.
%
Incoherent scattering adheres to the scale separation conjecture.

Following the observation of coherent scattering,
we want to emphasize the nuanced use of the terms \emph{source effects} and \emph{path effects}, which we will use consistently in the remainder of this article.
While our goal is to disentangle source effects from path variations, in reality our method more accurately isolates coherent features from within receiver groups that may encompass coherent scattering, 
e.g. surface-reflected arrivals pP and sP, source-side reverberations, and intrinsic attenuation of high frequencies, in addition to the primary source information.
In other words, coherent scattering is path independent and cannot be effectively disentangled from the source information using our methods due to the violation of scale separation (Fig.~\ref{fig:intro_scale}).
We synonymously use the term \emph{path effects} to loosely refer to incoherent scattering.

\section{Training Data}
\label{sec:training_data}

\begin{table}
\centering
\input{new_table.tex}
\caption{The SymVAE model was trained on displacement seismograms from these deep-focus earthquake events. The table provides the codenames of the earthquakes referenced throughout the paper.}
\label{tab:eq_details}
\end{table}
 
This section outlines the methodology used to construct the dataset for training SymVAE.
Broadband displacement seismograms of selected deep earthquakes (listed in Tab.~\ref{tab:eq_details}) were retrieved from the IRIS data center using the ObspyDMT package~\citep{hosseini2017obspydmt}. The seismograms were then instrument-corrected and manually inspected to remove outliers. Subsequently, standard pre-processing techniques, such as band-pass filtering, resampling, and standardization, were applied to the remaining seismograms.
During this study, our analysis focused on the time window of $\pm200\,$s centered around the IASP91~\citep{kennetttraveltimes} P or pP arrivals of the seismograms. 
The seismograms within the selected windows undergo polarity correction, ensuring uniform polarity across all seismograms associated with a particular earthquake.
The length of the time window is not critical as long as it is sufficient to capture the source effects. 
Furthermore, the same analysis can be adapted to other time windows by identifying relevant arrivals.

In order to utilize SymVAE for each earthquake, the seismograms were divided into multiple groups based on the idea that the receivers in each group capture identical apparent source function, which is sensitive to the direction (represented by $\gamma$) in which the waves depart the source region.
To aid in this analysis, we used a pixelated focal sphere, as implemented in HEALPIX, which divided the spherical surface into pixels of equal surface area with an angular resolution of $14\degree$.
As depicted in Fig.~\ref{fig:pixels}, the pixels were numbered and arranged in order to increase the azimuthal angle while maintaining constant polar angles, with higher pixel indices corresponding to higher values of the polar angle $\phi$. This distribution allowed us to easily examine the azimuthal dependence of the apparent source function by selecting pixels with constant polar angles.
The seismograms were grouped by tracing rays from the source region to all the receivers using the IASP91 model and assigning each receiver to a pixel index based on the intersecting pixel of the emanating ray from the source. For instance, Fig.~\ref{fig:pixels} displays the grouped receivers for the Bonin earthquake (with code bon1 in Tab.~\ref{tab:eq_details}). This grouping strategy was applied to all earthquakes in the dataset, ensuring that seismograms from different earthquakes were grouped separately. On average, 10 groups per earthquake were obtained, although this depended on data availability.
Furthermore, it should be noted that during the grouping and subsequent analysis, all available components of the displacement field were processed independently.
Following the grouping process, we discarded any groups that contained fewer than 15 seismograms for any displacement component. To ensure that coherent signals in each group of seismograms can be attributed exclusively to the earthquake source, it is vital that the path factors differ among the receivers within a group. When a group has only a few receivers positioned closely together, there is a risk that local geological features or nearby seismic sources might exert influences that are not sufficiently distinct between the receivers in that group.
%

%


Our framework allows for the training and analysis of a large number of earthquakes, utilizing all available grouped seismograms to form a set of training datapoints
$\{\Data{okt}{k}\}$. Each datapoint $\Data{okt}{k}$ is filled with time-windowed seismograms corresponding to a label $\source{okt}{k}$, which contains the okt earthquake code and the pixel index $k$ of the focal sphere. The $j$th preprocessed seismogram in $\Data{okt}{k}$
is
denoted by the notation
\begin{eqnarray}
    \data{okt}{k}{okt}{j},
\end{eqnarray}
where the superscript indicates the effects caused by the path connecting the source to the receiver station.
The usefulness of this unconventional notation will become apparent when we introduce the generation of virtual seismograms later.
A datapoint comprises a collection of seismograms, such as those associated with the $k$th pixel of the okt earthquake, expressed by:
\begin{eqnarray}
    \Data{okt}{k} = \left[\data{okt}{k}{okt}{1},\,\data{okt}{k}{okt}{2},\,\data{okt}{k}{okt}{3},\,\cdots\right].
\end{eqnarray}
The subsequent section will focus on developing an effective representation of these training data points.

\begin{figure*}
\centering
\includegraphics[width=\textwidth]{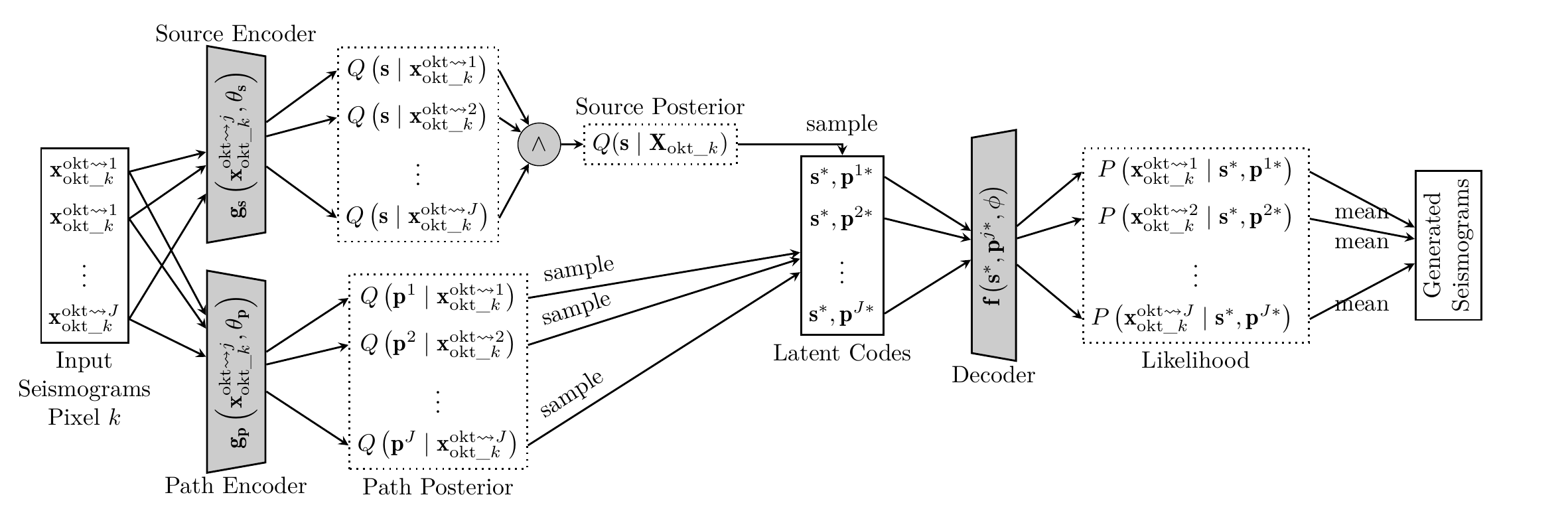}
\caption{
  \label{fig:network}
  Variational symmetric autoencoder architecture.
The source and path encoder networks address the inverse problem by estimating the posterior distributions.
The source encoder accumulates source information across all input seismograms linked to a particular focal sphere pixel. The decoder generates seismograms by executing forward modeling with the latent codes as input. SymVAE disentangles the source information from the path information within the latent space.
%
%
}
\end{figure*}

\section{Variational Symmetric Autoencoder}
A comprehensive explanation of variational symmetric autoencoders is provided by \cite{bharadwaj2024extractingcoherentseismicwavefield}; in this text, we will focus on the essential details and an intuitive explanation.
Variational autoencoder~\citep[VAE]{doersch2016tutorial} is designed to optimize networks that simultaneously solve the inverse or inference problem and the forward problem. The encoder addresses the inverse problem by estimating the posterior distribution of the latent variables given the observed seismograms.
In this context, latent variables correspond to the hidden features responsible for the generation of seismograms. 
Hidden features could consist of the earthquake's source-time function, wave attenuation, dispersion properties, scattering attributes, and various other forms of noise.
For a given seismogram, the encoder is tasked with identifying a set of features, whether they pertain to source or path effects, such that the seismogram can be
represented using these features.
The decoder, on the other hand, solves the forward problem by estimating the likelihood distribution from the latent codes; that is, it generates the seismograms given the features identified by the encoder. 
Optimizing the encoder and decoder parameters will ensure that not only the latent representations accurately capture significant features of the seismograms but also that the input seismograms are reconstructed without any loss of information.
Representation learning has gained prominence as a method of identifying useful representations tailored to the specific task being addressed.
Without constraints on the architectures of the encoder and decoder, obtaining a meaningful or interpretable representation becomes challenging. In other words, basic representation learning is problematic since numerous encoder/decoder configurations with varying latent variable settings can fulfill the training requirements.
Our primary aim is to obtain a meaningful representation of seismograms, where the latent features linked to the source effects are disentangled. We demonstrate that the principle of scale separation, which asserts that the source information changes at a slower rate, will offer essential constraints to achieve this disentanglement.
In summary,
to understand the displacement field caused by earthquakes, it is necessary to convert the representation of the measured displacement from its original time-position coordinates to the source-path coordinates. This conversion process involves non-linear transformations, which we aim to learn in an unsupervised manner through symmetric variational autoencoders.

We will proceed to elucidate the concept of disentanglement by examining a group of seismograms denoted as $\Data{okt}{k}$, which correspond to a particular pixel from an earthquake. The SymVAE encoder produces a latent code to represent $\Data{okt}{k}$. The encoder architecture is constrained to ensure that certain dimensions of this latent code uniquely capture coherent source information that is consistent across all seismograms in the group.
In other words,
the latent variable is partitioned into the following components.
\begin{itemize}
\item {\bf Source component ($\s$):} captures the coherent source information accumulated from all the seismograms of $\Data{okt}{k}$.
\item {\bf Path component ($\p^j$):} represents the effects of the path scattering and other noises unique to the $j$th seismogram within the group $\Data{okt}{k}$. 
\end{itemize}
SymVAE infers the random variables $\s$ and $\p^j$ independently via two separate encoder networks. Essentially, as described below, these networks are trained to produce posterior distribution parameters, with $Q(\s \mid \Data{okt}{k})$ corresponding to the source component and $Q(\p^j \mid \data{okt}{k}{okt}{j})$ pertaining to the path component.

To better understand how $\s$ is inferred, we focus on the role of the coherent source encoder in the SymVAE framework. This encoder directs the disentanglement procedure by being restricted to identify only the characteristics common to all seismograms within the group $\Data{okt}{k}$.
   To derive $Q(\s \mid \Data{okt}{k})$, one begins by employing a \emph{coherent source encoder} network $\g_s$ parameterized by $\theta_\s$. This network processes each individual seismogram $\data{okt}{k}{okt}{j}$ to produce the posterior distribution: $Q(\s \mid \x_j^k, \theta_\s)$. 
The next step involves identifying each seismogram as an independent observation containing source information, thereby combining the source information across all seismograms.
   %
   As shown in the Fig.~\ref{fig:network},
      we use the notion of \emph{conjunction of information states}~\citep{tarantola2005inverse} to combine source information across all seismograms. That means that the accumulated posterior is given by \begin{equation}
    Q(\s \mid \Data{eq}{k}, \theta_\s) = Q(\s \mid \data{eq}{k}{eq}{1}, \theta_\s)\land Q(\s \mid \data{eq}{k}{eq}{2}, \theta_\s)\land\cdots \propto \prod_{j}^{} Q(\s \mid \data{eq}{k}{eq}{j}, \theta_\s),
    \end{equation}
    where $\land$ denotes the conjuction.
    The key observation is that this process of accumulation of source information is
\emph{symmetric} with respect to the ordering of seismograms. Note that the precision of the source information increases as more seismograms are combined.
    %
   
    %

%


To minimize reconstruction loss, it is essential to encode all the features observed in the seismograms.
After investigating how the coherent source encoder gathers source information, we shift our attention to analyzing the extraction of the remaining features from the seismograms. These features are primarily path effects, unless the seismograms are noisy, in which case they will arise from both path effects and noise. 
Deriving path effects from seismograms is comparatively straightforward. 
For each seismogram indexed by $j$, a \emph{path encoder} network $\g_{\nui}$ parameterized by $\theta_{\nui}$ extracts these unique effects from each seismogram and produces $Q(\p^j \mid \data{eq}{k}{eq}{j}, \theta_{\p})$.

    %
    A decoder $\f$ with parameters $\phi$ is trained for forward modeling using samples from the posterior distributions of both $\s$ and $\p^j$ produced by the encoders.
     Denoting $\s^{*}$ and $\p^{j*}$ as samples drawn from the posterior distributions $Q(\s \mid \Data{okt}{k})$ and $Q(\p^j \mid \data{okt}{k}{okt}{j})$, respectively, the decoder then outputs the likelihood distribution $P(\data{okt}{k}{okt}{j}\mid\s^{*}, \p^{j*}, \phi)$, where its mean is given by $\f(\s^{*}, \p^{j*}, \phi)$.
     As shown in Fig.~\ref{fig:network}, the likelihood distribution for each seismogram in $\Data{okt}{k}$ is generated independently.
Using the sampling approach, we capture the probabilistic nature of the data generation process and enable the generation of synthetic seismograms by drawing random samples from the learned posterior distribution.

    The training procedure focuses on adjusting the parameters ($\theta_{\s}$, $\theta_{\p}$ and $\phi$) of both the encoders and the decoder to enhance the likelihood of generating the observed seismograms.
    Put differently, 
   by gradually fine-tuning these parameters in the course of training, SymVAE's encoder develops the capability to convert input seismograms into a latent space representation of the source, path, and noise effects, while the decoder learns to generate seismograms that align with the observed seismograms.
   One potential application of SymVAE is to use its decoder to generate virtual seismograms where the source effects are derived from one earthquake and the path effects from another. Appendix~\ref{app:virtual} provides a detailed explanation of this procedure and evaluates the accuracy of the virtual seismograms, using an experiment involving collocated earthquake sources. 
 This experiment demonstrates that the SymVAE network has been effectively trained. 
  We will now concentrate on harnessing the generative potential of SymVAE to explore whether it is feasible to generate seismograms with reduced noise and scattering effects.

\subsection{Generating Source Time Functions}
Upon training SymVAE, we have gained the ability to generate an ensemble of virtual seismograms that retain the same source characteristics while exhibiting differences in noise levels and path propagation effects.
To illustrate, let us look at the procedure for producing virtual seismograms using the source information from the $k$th pixel of the okt earthquake. We need a set of path codes to generate these virtual seismograms via the SymVAE decoder for this earthquake. To derive the path codes, we randomly choose another earthquake (denoted by code eq) and one of its pixels (indexed as $l$). We then sample from the associated posterior distribution $Q(\p^j \mid \data{eq}{l}{eq}{j})$. Once the path codes are obtained, for each sampled path code denoted by $\m$, the virtual seismogram generated is expressed as
\begin{equation} 
  \virt{okt}{k}{\m} = \f(\smean{okt}{k}, \m, \phi), 
  \label{eqn:virtm}
\end{equation} 
where $\smean{okt}{k}$ signifies the average of the source posterior distribution $Q(\s \mid \Data{okt}{k}, \theta_\s)$.
Intuitively, keep in mind that in the course of forward modeling, the decoder $\f$ discerns the source signature of the source code $\smean{okt}{k}$ and determines the noise and path effects to apply according to the input code $\m$.
For various samples of $\m$, we are able to generate seismograms that demonstrate different path effects. 

With an ensemble of virtual seismograms at our disposal, the next step is to pinpoint those that are essential to deduce the characteristics of the source.
In other words, 
although many of the resulting virtual seismograms may be noisy and impractical, our goal is to identify a virtual seismogram with the most accurate source details, specifically one that exhibits minimal noise and path effects.
 According to \cite{bharadwaj2024extractingcoherentseismicwavefield},
 the Kullback-Leibler (KL) divergence:
\begin{equation}
  \KLk(\virt{okt}{k}{\m}) = \KL{Q(\coh \mid \Data{okt}{k})}{Q(\coh \mid \virt{okt}{k}{\m})}
  \label{eqn:klopt}
  \end{equation}
can be used to identify virtual seismograms 
which have maximum source (coherent) information and minimal path effects.
Here, $Q(\coh \mid \virt{okt}{k}{\m})$ represents the source information in the virtual seismogram $\virt{okt}{k}{\m}$, while $Q(\coh \mid \Data{okt}{k})$ is the source information accumulated from all the seismograms associated with the $k$th pixel of the earthquake with the okt code.
The KL divergence measures how much one probability distribution differs from another, acting as a \emph{distance} that quantifies how much source information is lost when the real seismograms in $\Data{okt}{k}$ are replaced with the virtual seismogram $\virt{okt}{k}{\m}$.
In other words, a lower KL divergence $\KLk(\virt{okt}{k}{\m})$ implies better retention of the original source information within $\virt{okt}{k}{\m}$.

In practice, rather than exploring a set of virtual seismograms to identify the best one, we minimize the KL divergence in Eq.~\ref{eqn:klopt} by determining the optimal path code $\m$ using a gradient descent approach.
In this process, the KL objective gradient is computed to determine the necessary modifications of the path code to improve coherent source data while minimizing undesirable effects, with the path code being iteratively updated until it converges.
The task of identifying $\m$ resembles instructing the decoder network to produce a virtual seismogram, which contains an equivalent amount of source information as the encoder gathered using all the seismograms for a particular pixel.
We represent the optimal path code derived through gradient descent for the okt earthquake with $\m^{\text{opt}}_{\text{okt}}$. It is important to mention that this code is inherently influenced by the selected pixel (here, denoted as $k$), but we omit this dependency since we consistently select the pixel with the maximum number of seismograms. The rationale is that the source information obtained from such a pixel is more accurate compared to one with a lesser number of seismograms.
The virtual seismogram produced using the optimal path code $\m^{\text{opt}}_{\text{okt}}$ is termed the SymVAE generated source time function (STF) and is represented by 
\begin{equation} 
  \stf{okt}{k} = \f(\smean{okt}{k}, \m^{\text{opt}}_{\text{okt}}, \phi), 
  \label{eqn:stf1}
\end{equation} 
for the individual earthquake.

\subsection{Directivity Analysis}

\begin{figure}
    \centering
        \includegraphics[width=0.8\textwidth]{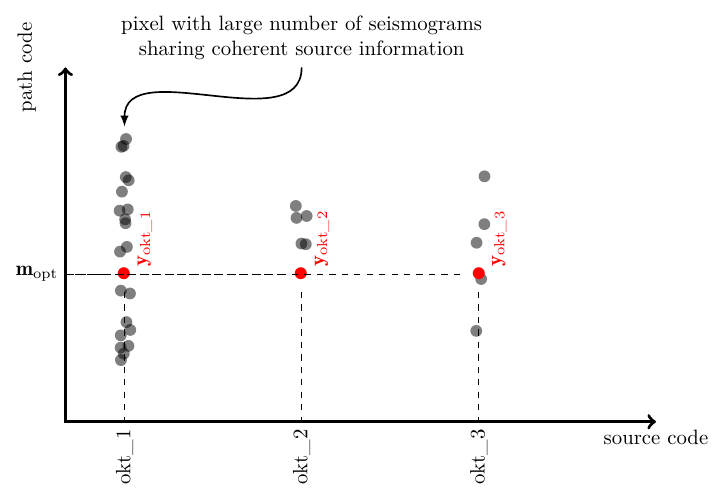}
                       \caption{
                        Each dot represents a seismogram plotted in the latent space of SymVAE. Gray dots correspond to measured seismograms, while red dots represent virtual seismograms with minimal path effects as generated by the decoder.
             In this context for the okt earthquake, the pixel 1 is equipped with numerous receivers, unlike pixels 2 and 3 which have fewer receivers and demonstrate subpar performance in envelope stacking. To determine the optimal path code, $\m^{\text{opt}}$, we minimize eq.~\ref{eqn:klopt} specifically for the first pixel, subsequently employing this code to create source time functions for pixels with fewer seismograms.
}\label{fig:redatuming1}
\end{figure}

To accurately analyze earthquake directivity and reliably assess the directional variations in source-time functions, it is vital to produce a source time function for each focal-sphere pixel that is devoid of path scattering effects and other noises.
Analyzing each pixel separately with methods such as envelope stacking leads to performance variations from one pixel to the next, influenced by the variability in seismogram quality and quantity at different pixels. Consequently, the directivity estimates derived from these techniques may be impacted by residual noise.
%
In contrast, SymVAE is trained on the seismograms of all pixels simultaneously ---
which means, it is capable of generating virtual seismograms by utilizing source data from one pixel, while extracting noise from high-SNR seismograms of a different pixel.
Intuitively, once SymVAE acquires the skill to mitigate noises for an individual pixel, this knowledge can be applied to effectively reduce noises across all pixels uniformly.
That is, after estimating the optimal path code, it can be reused to minimize noises across all pixels, including those with a low signal-to-noise ratio or limited receiver count, where envelope stacking is less effective.
Again, 
in the case of the okt earthquake, once the optimal path code tailored to the pixel with the highest receiver count is determined by optimizing eq.~\ref{eqn:klopt}, this code can be leveraged to construct source time functions for the remaining pixels related to that earthquake. This is shown graphically in Fig.~\ref{fig:redatuming1}. We denote the STF collection for every pixel of the okt earthquake with the vector
\begin{eqnarray}
    \label{eqn:usvs1}
    \Stf{okt} = \left[\stf{okt}{1}, \stf{okt}{2},\, \cdots\right] = 
    \left[\f(\smean{okt}{1}, \m^{\text{opt}}_{\text{okt}}, \phi), \f(\smean{okt}{2}, \m^{\text{opt}}_{\text{okt}}, \phi),\,\cdots\right],
\end{eqnarray}
where $\smean{okt}{k}$ represents the coherent source information of $k$th pixel.
We note that $\m^{\text{opt}}_{\text{okt}}$ is earthquake specific, i.e., needs to be estimated individually for each earthquake.
In the next section, we will quantitatively assess the performance of SymVAE using a synthetic experiment.

\section{Synthetic Experiment}

\begin{figure}
    \centering
    \includegraphics[width=0.8\linewidth]{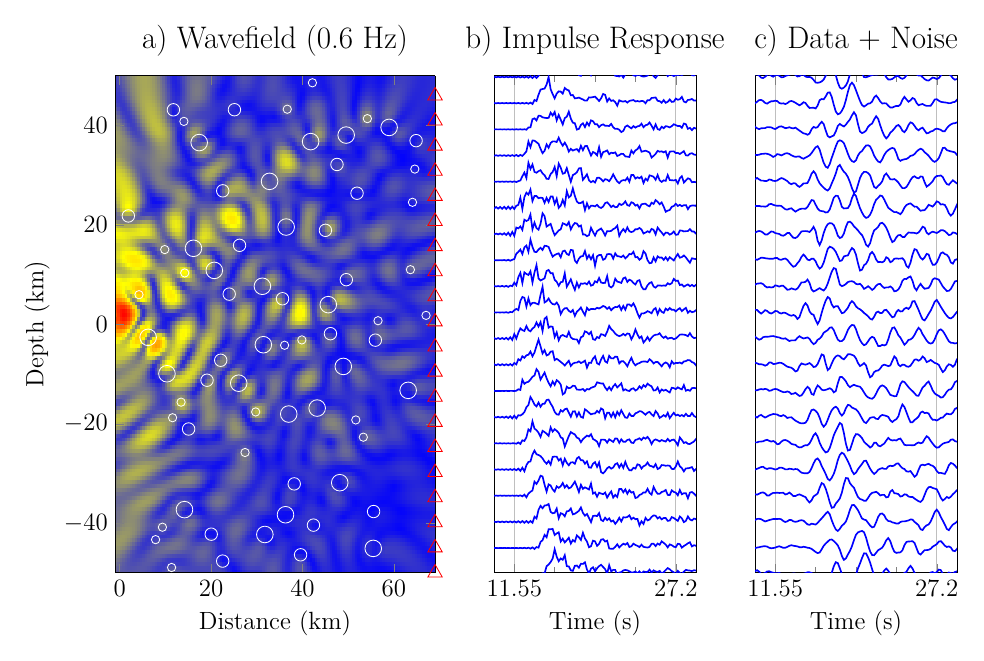}
    \caption{%
    a) Absolute value of the pressure wavefield in the frequency domain. A point source is located at the origin. Receivers are plotted as red triangles. Medium contains circular scatterers of varying radii.
    b) Normalized impulse response at selected receivers, highlighting scattering effects. 
    c) Synthetic pressure seismograms after convolving the response in (b) with an arbitrary source signature. These seismograms
    are used for training SymVAE.
    \label{fig:syn_setup}}
\end{figure}

\begin{figure}
    \centering
    \includegraphics[width=1\linewidth]{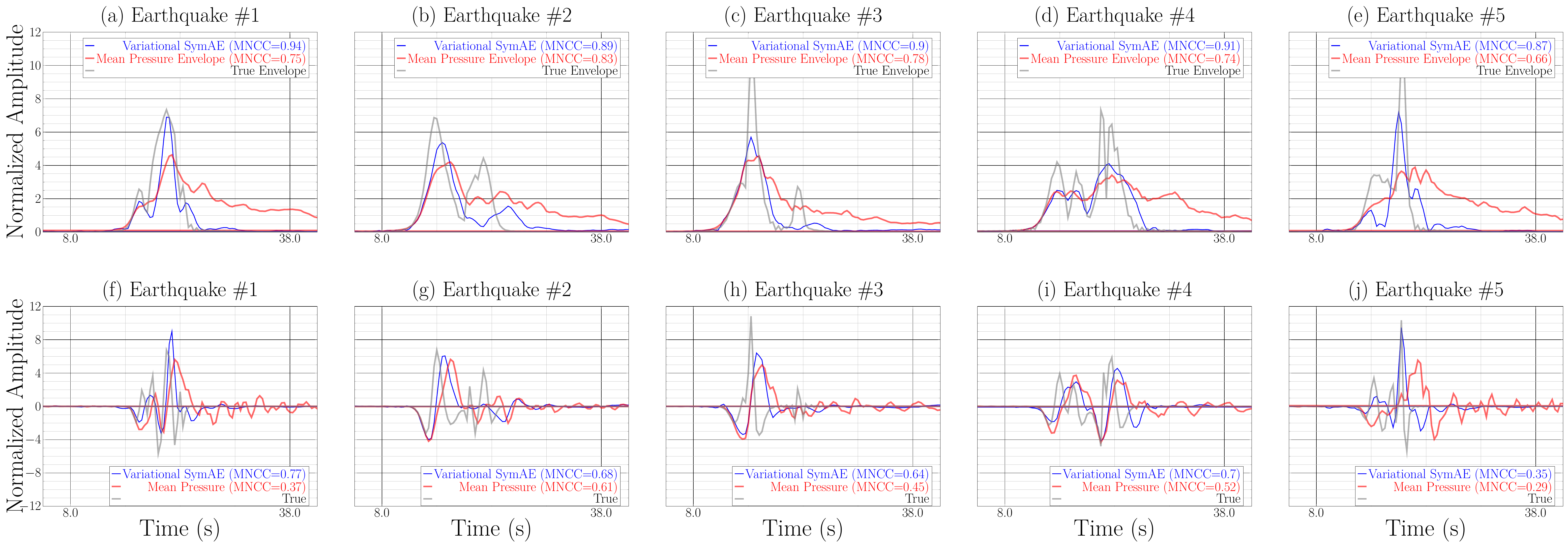}
    \caption{
SymVAE retrieval of STFs for synthetic earthquakes. Simple wavefield averaging exhibits poor resolution and residual scattering effects, as indicated by low maximum normalized cross-correlation (MNCC) values.
    \label{fig:syn_alleq_stf}}
\end{figure}

We performed a synthetic simulation using the \texttt{MultipleScattering.jl} library~\citep{Gower_2018, multiple_scattering_jl}, 
focusing on two-dimensional wave propagation in an attenuating acoustic medium containing circular scatterers of varying sizes. 
The background velocity of the medium is 5.8\,km/s, and the scatterers have a 30\% velocity contrast.
For simplicity, a point-earthquake source with an arbitrary signature is placed at the origin.
Future work may involve conducting synthetic tests with various rupture scenarios and linear elasticity.
The scattered pressure wavefield, shown in Fig.~\ref{fig:syn_setup}a, is sampled at 100 receivers
to train the network together with the data described in Section~\ref{sec:training_data}.
With complex scattering, as evident in the impulse response of Fig.~\ref{fig:syn_setup}b, we aim to use SymVAE and blindly extract the coherent source time function (STF) from the pressure seismograms in Fig.~\ref{fig:syn_setup}c.
Toward that end, the optimal path code $\m^{\text{opt}}_{\text{eq}}$ for the earthquake is determined by minimizing eq.~\ref{eqn:klopt}, then eq.~\ref{eqn:stf1} is applied to reconstruct the STF. 
In Fig.~\ref{fig:syn_alleq_stf}, we compare true STFs with SymVAE-derived STFs for five synthetic earthquakes. In general, SymVAE STFs appear smoother, indicating that loss of high-frequency information due to attenuation is irreversible.
%
%
As the source is isotropic, we have also performed wavefield stacking to average out the scattering effects and derive STFs.
We use maximum normalized cross-correlation (MNCC) as a robust metric to quantify optimal alignment and similarity between two signals.
The plots Figs.~\ref{fig:syn_alleq_stf}(f)--(j) show a higher MNCC between SymAE and true STFs as opposed to the wavefield stacking and true STFs. 
For real earthquakes, faulting complicates the wavefield stacking due to polarity differences, so we stack the envelope
(instantaneous amplitude computed via the Hilbert transform) of the pressure wavefield instead. Again, SymVAE exhibits a better correlation (see Figs.~\ref{fig:syn_alleq_stf}(a)--(e)) with true STF envelopes compared to stacking --- observe how residual scattering effects during stacking obscure the stopping phase.
This synthetic experiment implies that SymVAE STFs have minimal incoherent scattering effects, and the data with 30 earthquakes in Section~\ref{sec:training_data} were adequate for training SymVAE.

\section{Results for Deep Earthquakes}

In this section, we shall analyze the SymVAE STFs pertaining to various deep earthquakes. 
Here, the reconstruction of positive (due to absence of backslip) moment-rate functions utilizing band-limited far-field data is ill-posed because of loss of information.
Therefore, gaining insights from the synthetic experiment, we interpret SymVAE-derived STFs as band-limited moment-rate functions.
The spectra of these STFs can be further extrapolated to recover full-band moment-rate functions using prior information, which is not the focus of this paper.
Moreover, we only plot SymVAE STF envelopes for a straightforward comparison with stacked envelopes or traditional deconvolution results.

\begin{figure}
    \centering
    \includegraphics[width=1\linewidth]{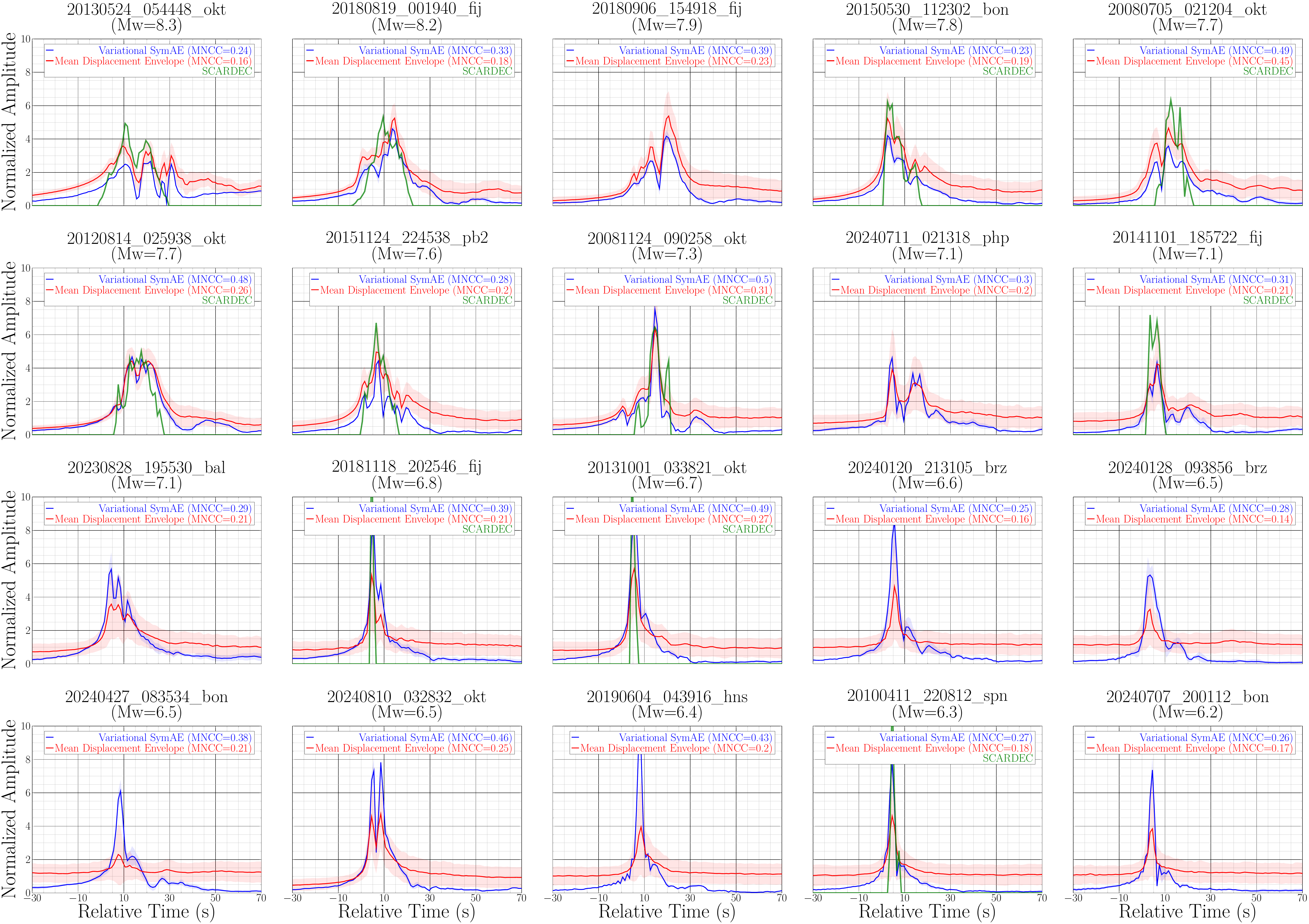}
    \caption{Comparison of blue SymVAE-derived P STF envelopes with 1) red optimized envelope stacking, using multiple nearby receivers for stacking, and 2) green non-array-based deconvolution method~\citep{vallee_scardec_2011}.
    For high-magnitude earthquakes, SymVAE yields similar, if not better outcomes, compared to envelope stacking (see Fig.~\ref{fig:decon_tree}). However, as the earthquake magnitude and signal-to-noise ratio (SNR) declines, SymVAE enhances weak secondary rupture episodes and offers superior resolution in contrast to the less effective envelope stacking results. 
    The maximum normalized cross-correlation (MNCC) with the observed displacement wavefield is calculated to demonstrate that SymVAE STFs provide a superior explanation of the wavefield compared to envelope stacking.
%
    \label{fig:alleq_stf}}
\end{figure}

 \begin{figure}
     \centering
     
     \begin{subfigure}[b]{0.225\textwidth}
         \centering
         \includegraphics[width=\textwidth]{plots_Mixed_Single_Multiple_Pixel_BestInterpreted_24Feb2025/compressed_pdfs/plotraw_20240427_083534_bon.jld2.pdf}
         \caption{Envelope Stacking of bon2}
         \label{fig:bon2_rawp}
     \end{subfigure}
     \begin{subfigure}[b]{0.225\textwidth}
         \centering
         \includegraphics[width=\textwidth]{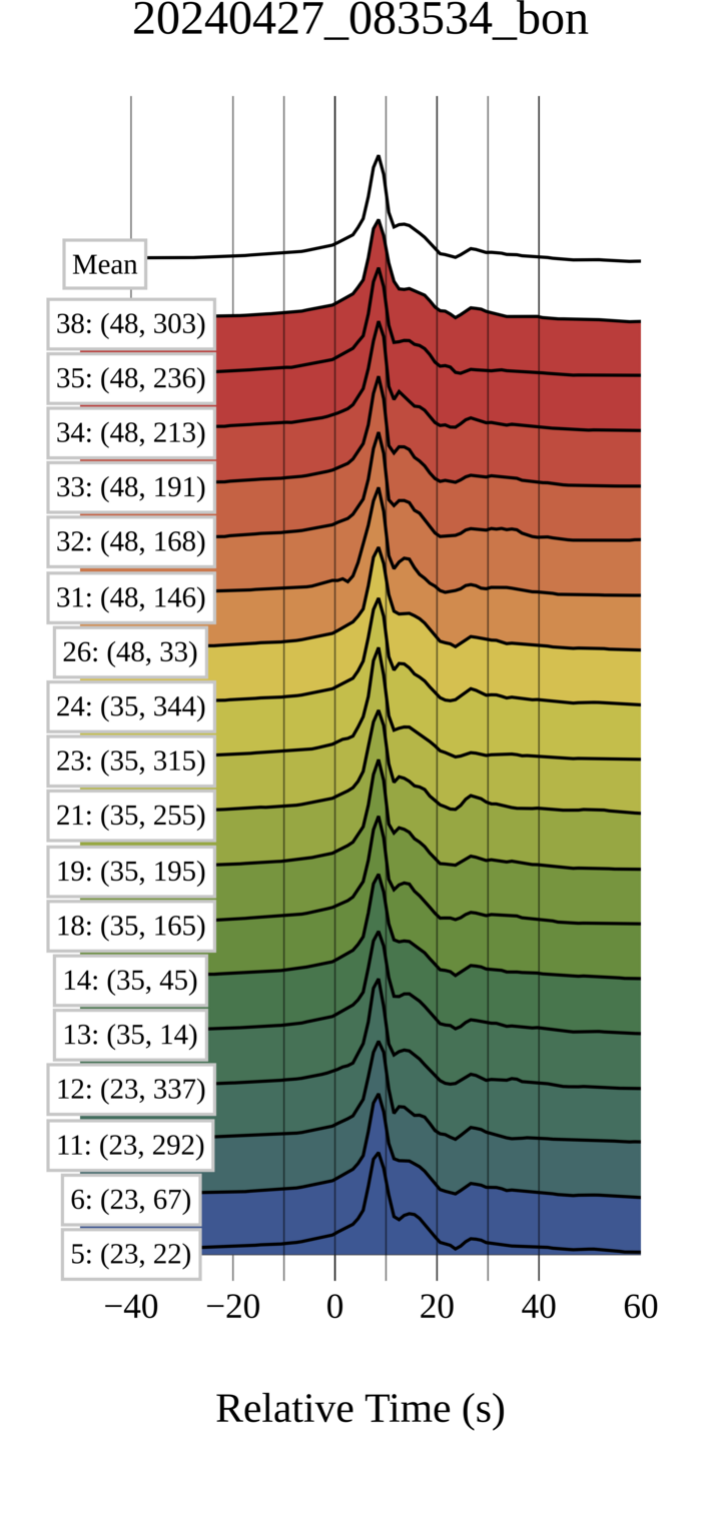}
         \caption{SymVAE STFs of bon2}
         \label{fig:bon2_usvsp}
     \end{subfigure}
     \begin{subfigure}[b]{0.225\textwidth}
         \centering
         \includegraphics[width=\textwidth]{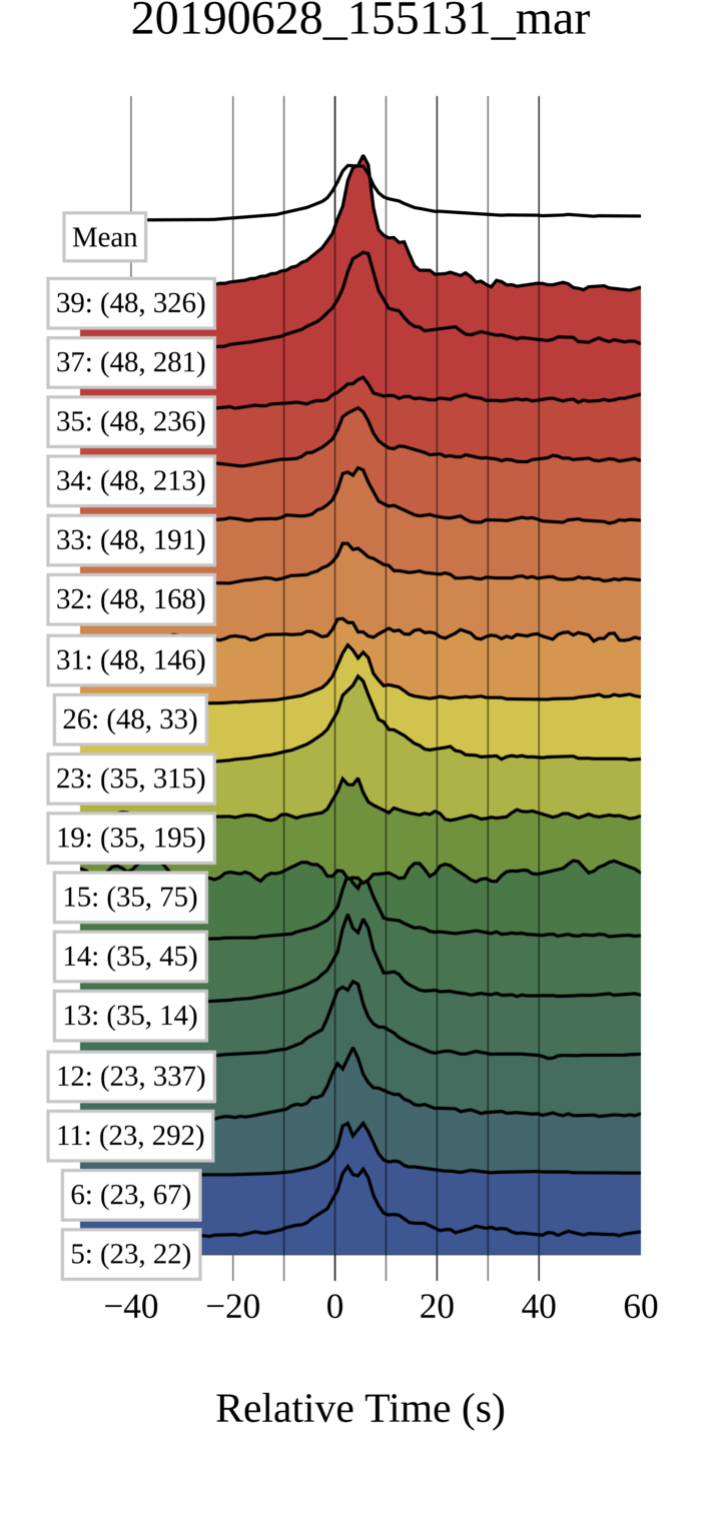}
         \caption{Envelope Stacking of mar}
         \label{fig:mar_rawp}
     \end{subfigure}
     \begin{subfigure}[b]{0.225\textwidth}
         \centering
         \includegraphics[width=\textwidth]{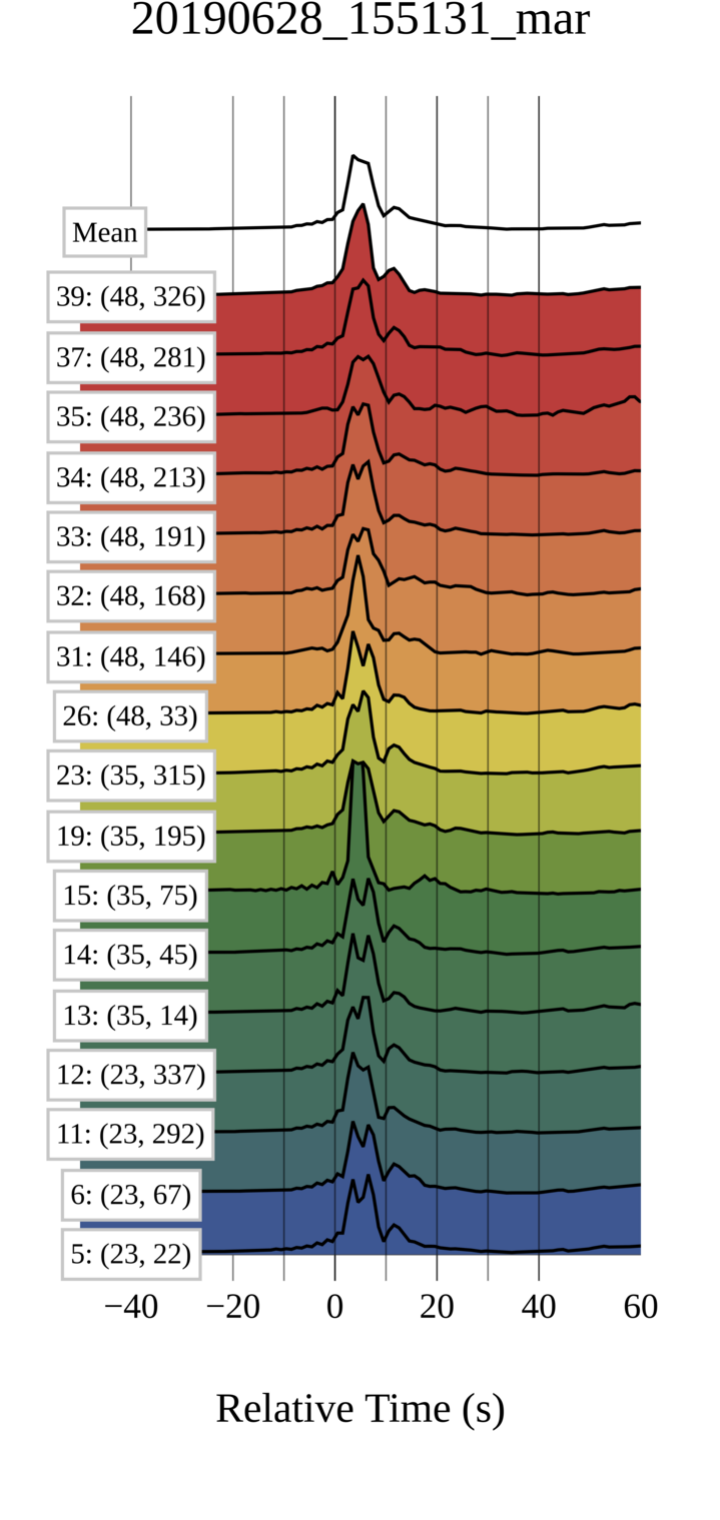}
         \caption{SymVAE STFs of mar}
         \label{fig:mar_usvsp}
     \end{subfigure}
        \caption{
    An examination of the P-wave directivity patterns of two earthquakes, both having magnitudes near 6.5 Mw, reveals that not much directivity is evident in either instance.
 In the case of envelope stacking, identifying the primary episode and its duration becomes challenging, especially in pixels with a limited number of receivers, as the SNR is low.
 Moreover, envelope stacking is unable to detect weaker secondary episode starting at approximately $12\,$s for both earthquakes.
        SymVAE STFs provide improved resolution across all pixels.
        %
        Pixels are labeled with their index followed by corresponding (polar angle, azimuthal angle) pair according to the pixelation shown in Fig.~\ref{fig:pixels}a.
        Interactive version of these plots can be accessed at \url{https://eq-SymAE.fly.dev/}.
        }
        \label{fig:bon2_mar}
\end{figure}

First, we will analyze the SymVAE STFs for a set of earthquakes, focusing only on the pixel that contains the highest number of receivers.
%
%
The envelope stacking is expected to perform optimally for these particular pixels --- as plotted in Fig.~\ref{fig:alleq_stf}, where the shaded bands represent the range with a standard deviation of one.
%
Since all seismograms undergo normalization before SymVAE training, it is not possible to retrieve the maximum amplitude of the STFs. Therefore, both envelopes and STFs are normalized prior to visualization.
For high-magnitude earthquakes in Fig.~\ref{fig:alleq_stf}, the stacking of envelopes provides a good approximation of the STFs for all earthquakes, since we choose pixels that contain seismograms of high quality and quantity.
%
However,
several instances in Fig.~\ref{fig:alleq_stf} clearly demonstrate the improvement in the source information derived from SymVAE.
\begin{enumerate}
    \item In most situations, SymVAE-generated STFs exhibit greater resolution compared to those obtained through envelope stacking. Examples include the Mw 7.8 okt earthquakes and the Mw 6.3 spn earthquake.
    \item SymVAE STFs show superior quality even in poor SNR conditions. The Mw 6.5 bon2 earthquake illustrates this situation.
    \item SymVAE enhances secondary episodes that are obscured by coda waves. These secondary episodes immediately follow a stronger one and may be challenging to identify.
 A prime example is the Mw 7.1 php earthquake, where the raw envelope becomes obscured between two major energy release episodes. However, SymVAE effectively identifies detailed patterns between these episodes. 
    \item For low-magnitude earthquakes, where weaker secondary episodes are generally difficult to detect via envelope stacking, SymVAE enhances these episodes despite being obscured by noise.
    Several instances are shown in the last two rows of Fig.~\ref{fig:alleq_stf}.
\end{enumerate}
We computed the average maximum normalized cross-correlation (MNCC) with the observed seismograms (within the selected pixel) and noticed that the SymVAE STFs correlate more with the seismograms compared to the envelope stacks. This suggests that the secondary episodes in the SymVAE STFs are physically real and not artifacts of the generative model. Furthermore, in Fig.~\ref{fig:alleq_stf}, we have compared some of the STFs with those derived using SCARDEC, a non-array-based deconvolution method~\citep{vallee_scardec_2011, vallee_new_2016-1}. There is a consensus on the duration of the main episode (see the Mw 7.1 fij3 earthquake). In addition, certain secondary episodes (see the Mw 7.3 okt4 earthquake) are detectable in these deconvolution outputs.

Our next step involves examining the slight variations related to the directivity in the STFs across the various pixels of a particular earthquake, as in Eq.~\ref{eqn:usvs1}.
We will first investigate the directivity of the P wave for two lower-magnitude earthquakes: the first, bon2, occurring near the Bonin Islands with a magnitude of Mw 6.5, and the second, mar, near the Mariana Islands with a magnitude of Mw 6.4.
In both cases,
the efficacy of envelope stacking was significantly dependent on the number of receivers connected to each pixel due to the low signal-to-noise ratio.
It is evident in Fig.~\ref{fig:bon2_mar}, that the variations in the stacked envelopes across different pixels are mainly influenced by the quality and quantity of the seismograms used.
In contrast, SymVAE-generated STFs demonstrate not only a distinct primary episode but also a weaker secondary episode (around 12\,s) for all pixels. It can be noticed that SymVAE can effectively extract coherent source information, even for pixels with limited receivers, compared to envelope stacking.
This ability to extract source information from noisy, low-magnitude deep earthquakes is crucial, especially since deconvolution methods are unstable under low-SNR conditions. 
For the analyzed earthquakes, the STFs exhibit minimal directivity effects, with only slight variations observed across the pixels. This indicates that the locations of the sources associated with various episodes are in close proximity. 
Furthermore, 
the minimal variation of the STFs across pixels also indicates that our findings are not influenced by path effects. Should path and noise have any influence, given the variability of these effects among various station networks, one would expect significant differences in results across different pixels.


We now examine the directivity patterns of P waves for two complex high-magnitude earthquakes, each comprising multiple episodes of nearly equal amplitude: 1) pb2, which pertains to the major deep-earthquake doublet that struck the Peru-Brazil border on November 24, 2015, as analyzed by ~\cite{doi:10.1126/sciadv.1600581}, and 2) okt7 near the Sea of Okhotsk with magnitude of 7.3.
SymVAE-derived STFs once again offer superior resolution compared to envelope stacking, despite
the fact that envelope stacking is expected to perform better in scenarios with high SNR. In the case of the pb2 earthquake, at least five rupture episodes have been resolved. However, no significant directivity was observed for this earthquake, again indicating that the source locations of the episodes are close by (low fault length). Considering the complexity of this earthquake, the SCARDEC traditional deconvolution method was unable to identify all the rupture episodes, as illustrated in Fig.~\ref{fig:alleq_stf} and discussed in Fig.~\ref{fig:decon_tree}.
%
%
For the okt7 earthquake, notable directivity is observed:
STFs associated with the azimuths between $22\degree$ and $56\degree$ appear compressed, indicating that the rupture is propagating in that specific direction, while
the opposite STFs connected to the azimuths spanning from $236\degree$ to $285\degree$ appear stretched. 
Recall that directivity refers to the phenomenon in which the direction of rupture propagation is characterized by shorter time intervals between successive episodes (higher frequency), while the opposite direction experiences longer intervals (lower frequency) between rupture episodes.
Finally, note that SymVAE reveals the presence of at least three distinct episodes for okt7 that envelope stacking does not detect.

 \begin{figure}
     \centering
     \begin{subfigure}[b]{0.225\textwidth}
         \centering
         \includegraphics[width=\textwidth]{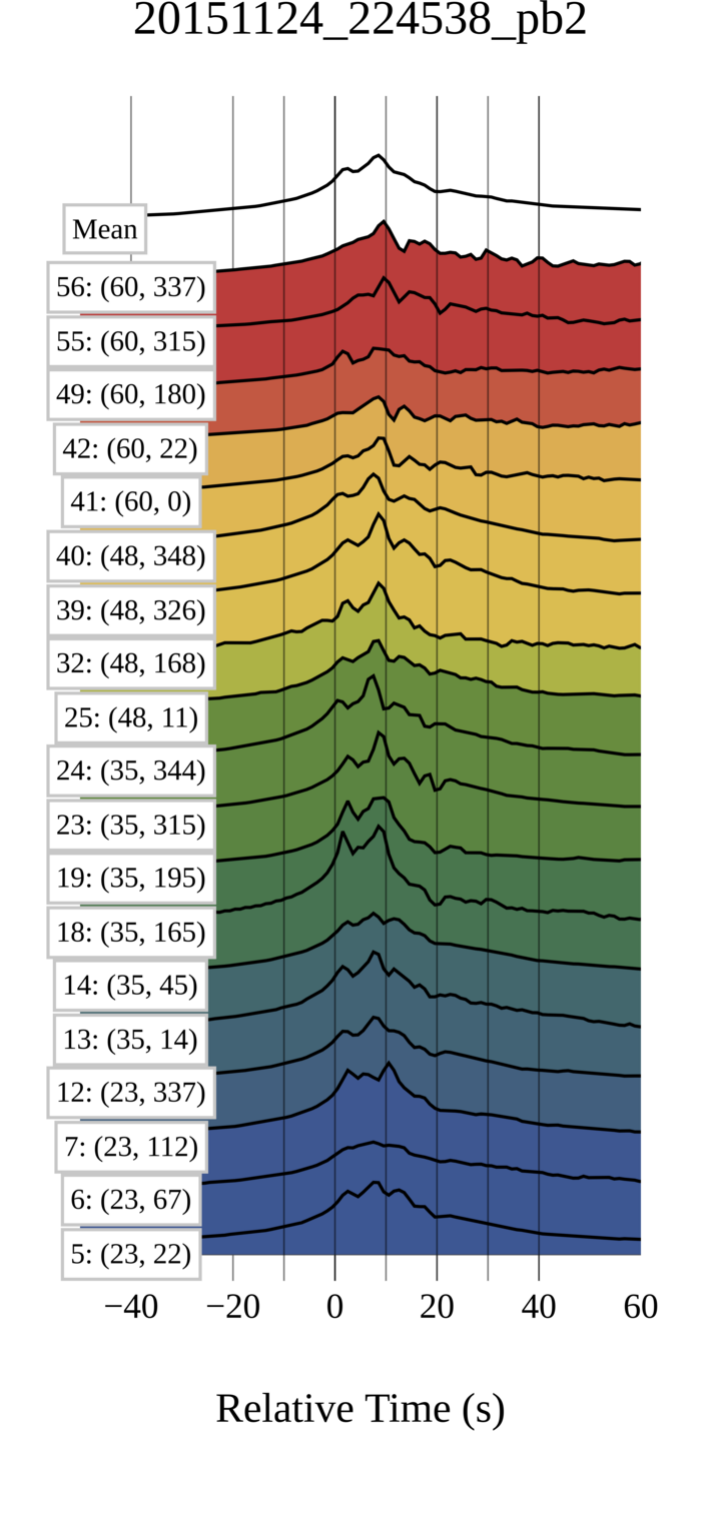}
         \caption{Envelope Stacking of pb2}
         \label{fig:pb2_rawp}
     \end{subfigure}
     \begin{subfigure}[b]{0.225\textwidth}
         \centering
         \includegraphics[width=\textwidth]{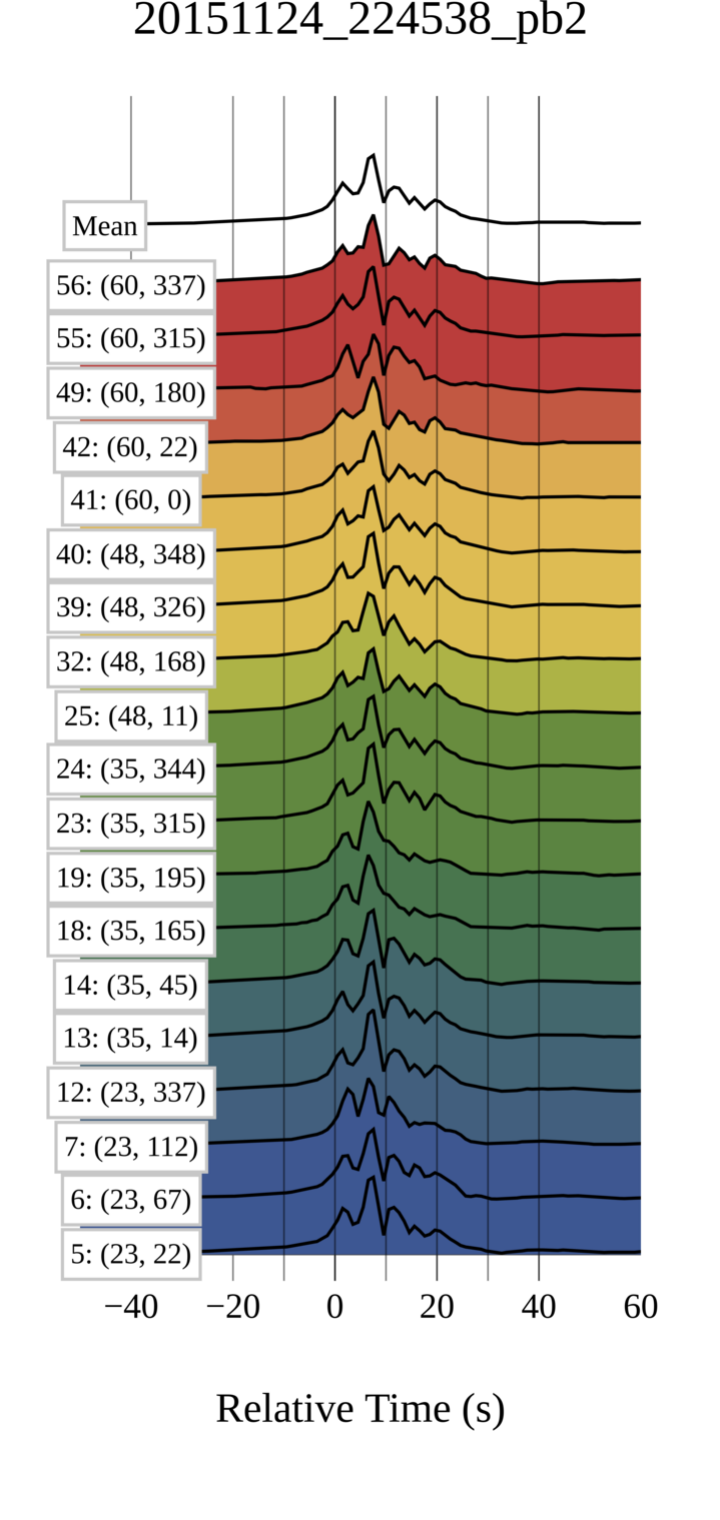}
         \caption{SymVAE STFs of pb2}
         \label{fig:pb2_usvsp}
     \end{subfigure}
     \begin{subfigure}[b]{0.225\textwidth}
         \centering
         \includegraphics[width=\textwidth]{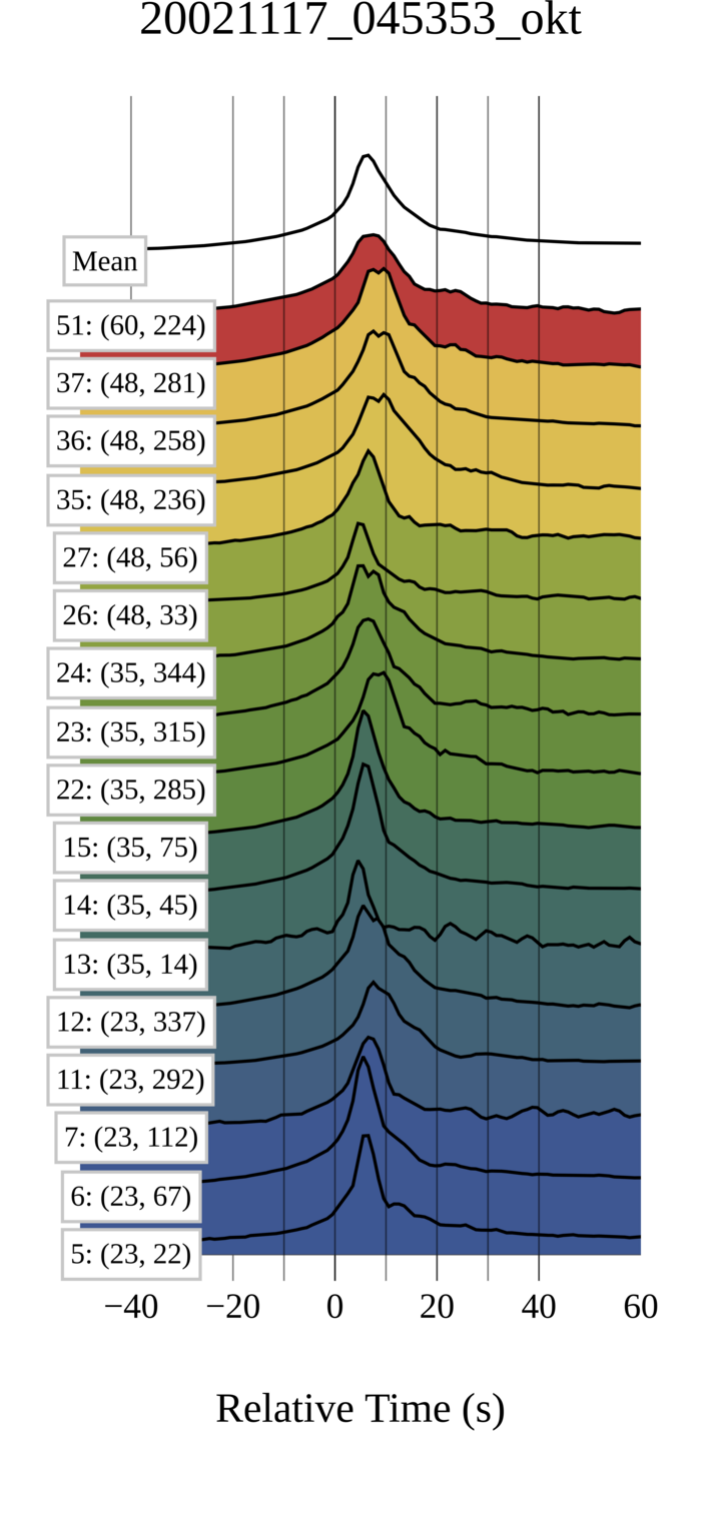}
         \caption{Envelope Stacking of okt7}
         \label{fig:okt7_rawp}
     \end{subfigure}
     \begin{subfigure}[b]{0.225\textwidth}
         \centering
         \includegraphics[width=\textwidth]{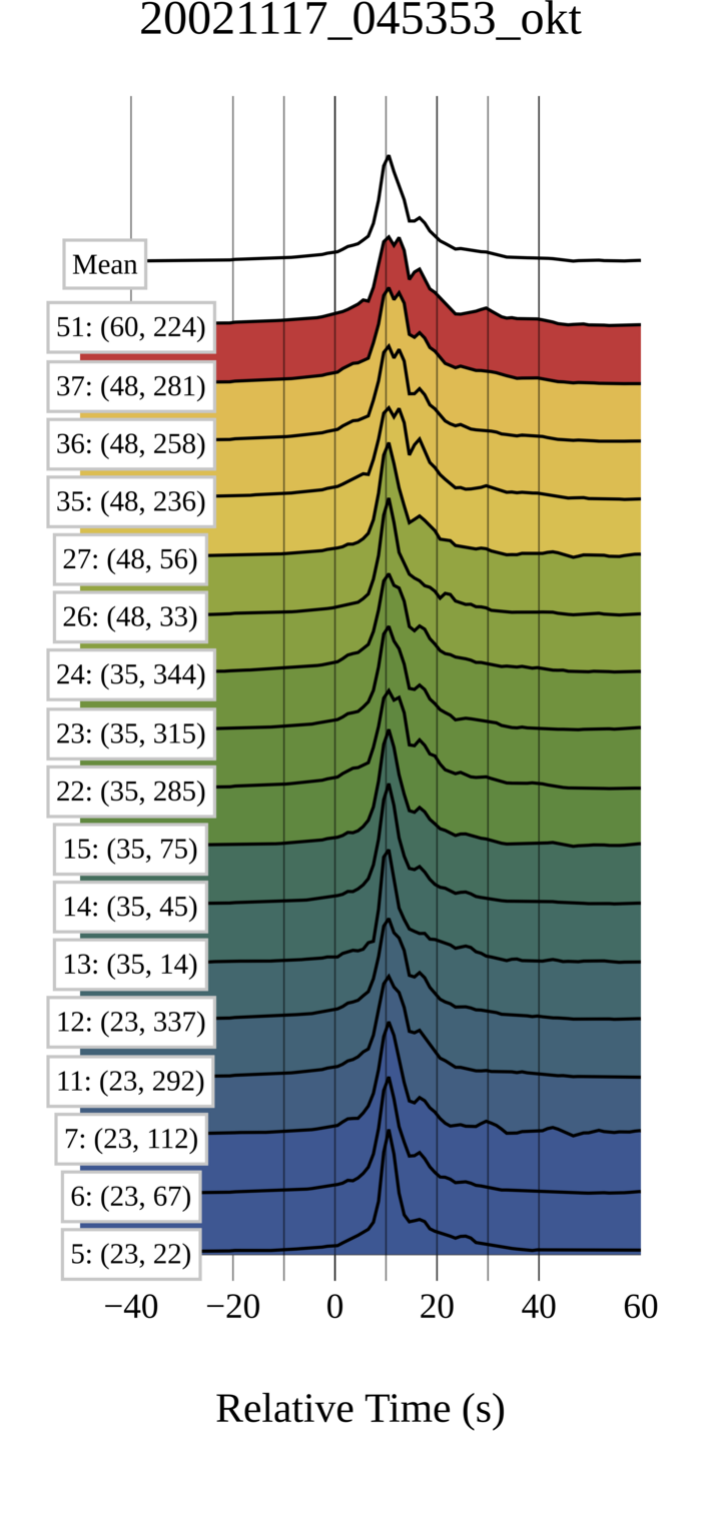}
         \caption{SymVAE STFs of okt7}
         \label{fig:okt7_usvsp}
     \end{subfigure}
        \caption{Analogous to Fig.~\ref{fig:bon2_mar}, this analysis focuses on larger magnitude earthquakes, which have a high signal-to-noise ratio and exhibit multiple episodes. The STFs obtained using SymVAE once more provide better resolution compared to envelope stacking.
    For the Mw 7.6 pb2 earthquake comprising a minimum of 5 episodes, minimal directivity effects are observed.
        In the case of the Mw 7.3 okt7 earthquake, significant directivity emerges where the STFs linked to pixels with azimuths ranging from $236\degree$ to $285\degree$ seem elongated, 
        compared to azimuths ranging from $22\degree$ to $56\degree$.
        More importantly, SymVAE reveals that this earthquakes is made 
        of at least three episodes that could not be interpreted through envelope stacking. Interactive version of these plots can be accessed at \url{https://eq-SymAE.fly.dev/}.
       }
        \label{fig:pb2_okt7}
\end{figure}


%

The high-resolution SymVAE STFs generated in this study have revealed critical information on the nature of deep-focus earthquakes. Our analysis reveals that these earthquakes are characterized by fragmented seismic moment release, occurring in multiple short episodes rather than continuous (long-duration) events. Specifically, none of the individual episodes we analyzed exceeded the duration of $5$--$10$ s. This fragmentation is observed in both low and high magnitude earthquakes, with lower magnitude events (Mw 6.0 to 7.0) comprising 2 to 3 episodes, and higher magnitude events (Mw > 7.0) consisting of 4 to 10 episodes. These findings support the multi-mechanism hypothesis~\citep{zhan2020mechanisms}, suggesting that deep-focus earthquakes are governed by a combination of processes, including metastable transformational olivine faulting, dehydration embrittlement, and thermal runaway. Our results provide compelling evidence that deep-focus earthquakes may initiate through one mechanism and propagate via another. 
Previous studies have reached similar conclusions about large-magnitude deep-focus earthquakes. For example, \cite{Chen1996} and colleagues identified seismic moment release bursts that occur in multiple episodes during large deep-focus earthquakes.
Furthermore, a study by \cite{Chen2015} suggested that the source processes of large deep-focus earthquakes can be best interpreted by cascading failures of shear thermal instabilities in pre-existing weak zones.
Finally, the results for all the earthquakes in Tab.~\ref{tab:eq_details} are available in \url{https://eq-SymAE.fly.dev/}, which serves as a portal to visualize the STFs generated by the symmetric autoencoders.

\section{Sanity Checks}

We proceed to conduct sanity checks to verify the source functions generated with SymVAE. These checks are crucial because the efficacy of a generative models such as SymVAE can only be evaluated on the basis of criteria related to the underlying physical phenomena.
The loss function minimized during the training process does not necessarily reflect the physical validity of the generated STFs, and the accuracy of them. 
In other words, minimizing the autoencoding loss during the training process does not
necessarily improve the accuracy of the generated STFs. 
The generated STFs are
merely interpolations of the measured seismograms in high dimensions.

\subsection{Comparison with envelope stacking}
To verify the accuracy of the SymVAE-generated STFs, we performed a comparison with envelope stacking in all the findings reported in this article. Envelope stacking is most effective when dealing with
high-quality pixels containing the maximum number of receivers.
As depicted in Fig.~\ref{fig:alleq_stf}, in this scenario, the pattern of rise and the number of episodes in the SymVAE extracted STFs roughly match those of the raw envelopes.

%
%

%


\subsection{Comparison between P and pP seismograms}
\begin{figure}
     \centering
     \begin{subfigure}[b]{0.225\textwidth}
         \centering
         \includegraphics[width=\textwidth]{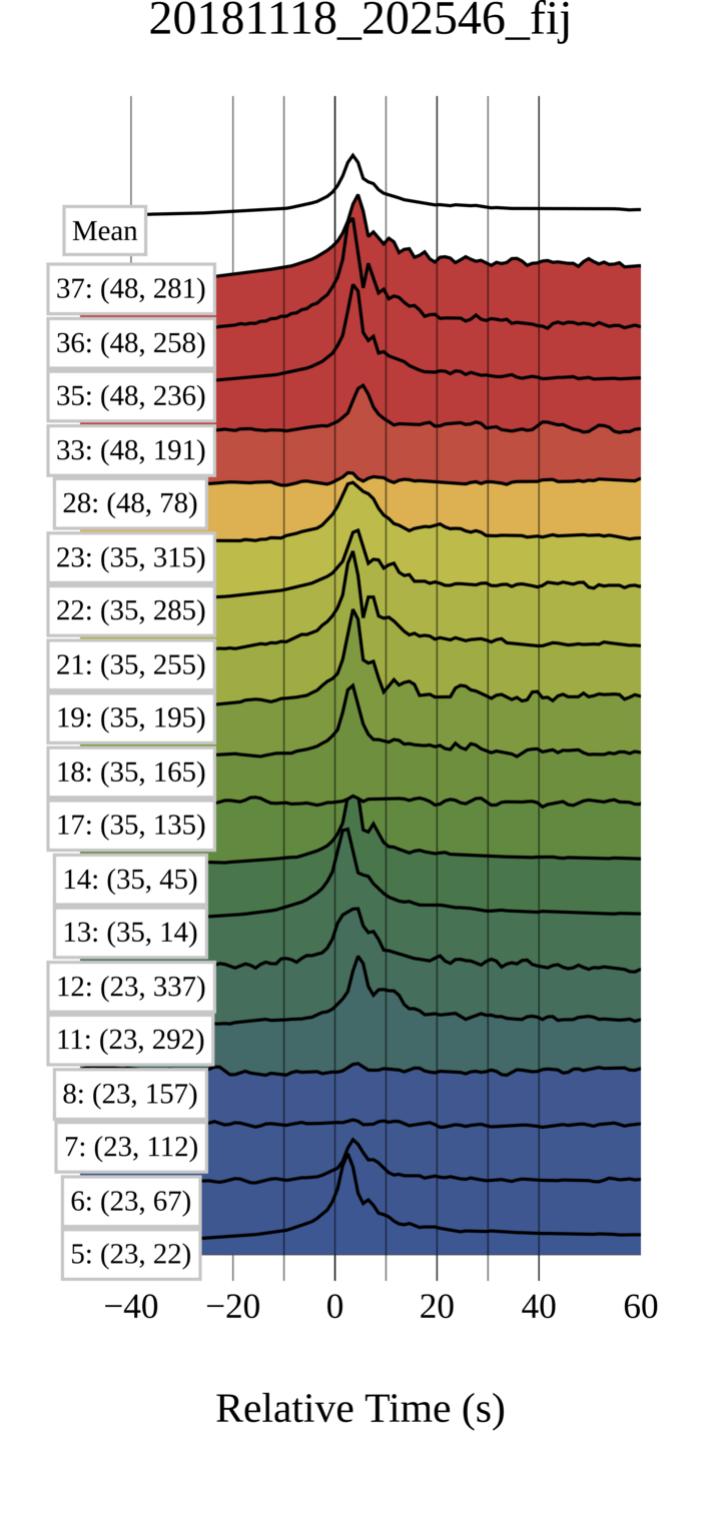}
         \caption{P Envelope Stacking}
         \label{fig:2002_okt_rawp}
     \end{subfigure}
     \begin{subfigure}[b]{0.225\textwidth}
         \centering
         \includegraphics[width=\textwidth]{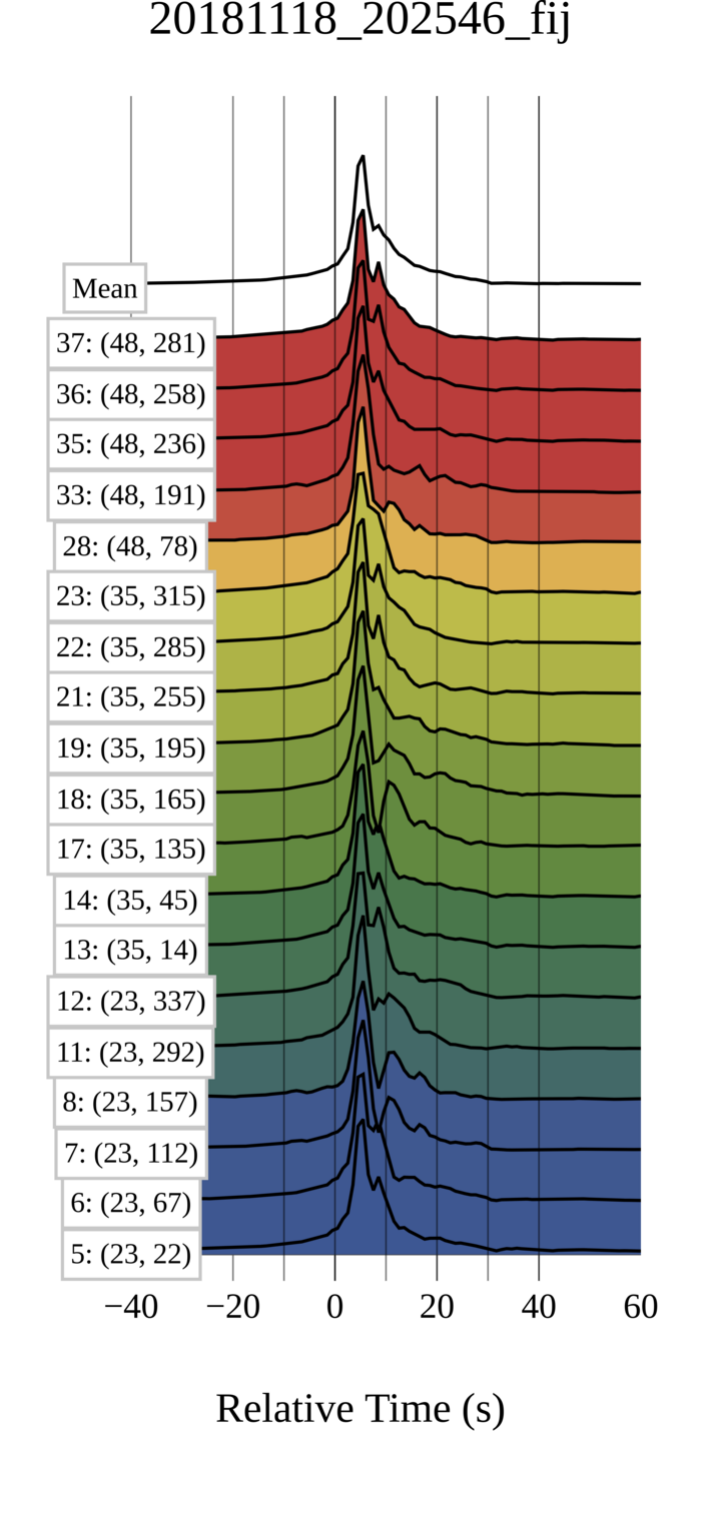}
         \caption{SymVAE P STFs}
         \label{fig:2002_okt_usvsp}
     \end{subfigure}
     \begin{subfigure}[b]{0.225\textwidth}
         \centering
         \includegraphics[width=\textwidth]{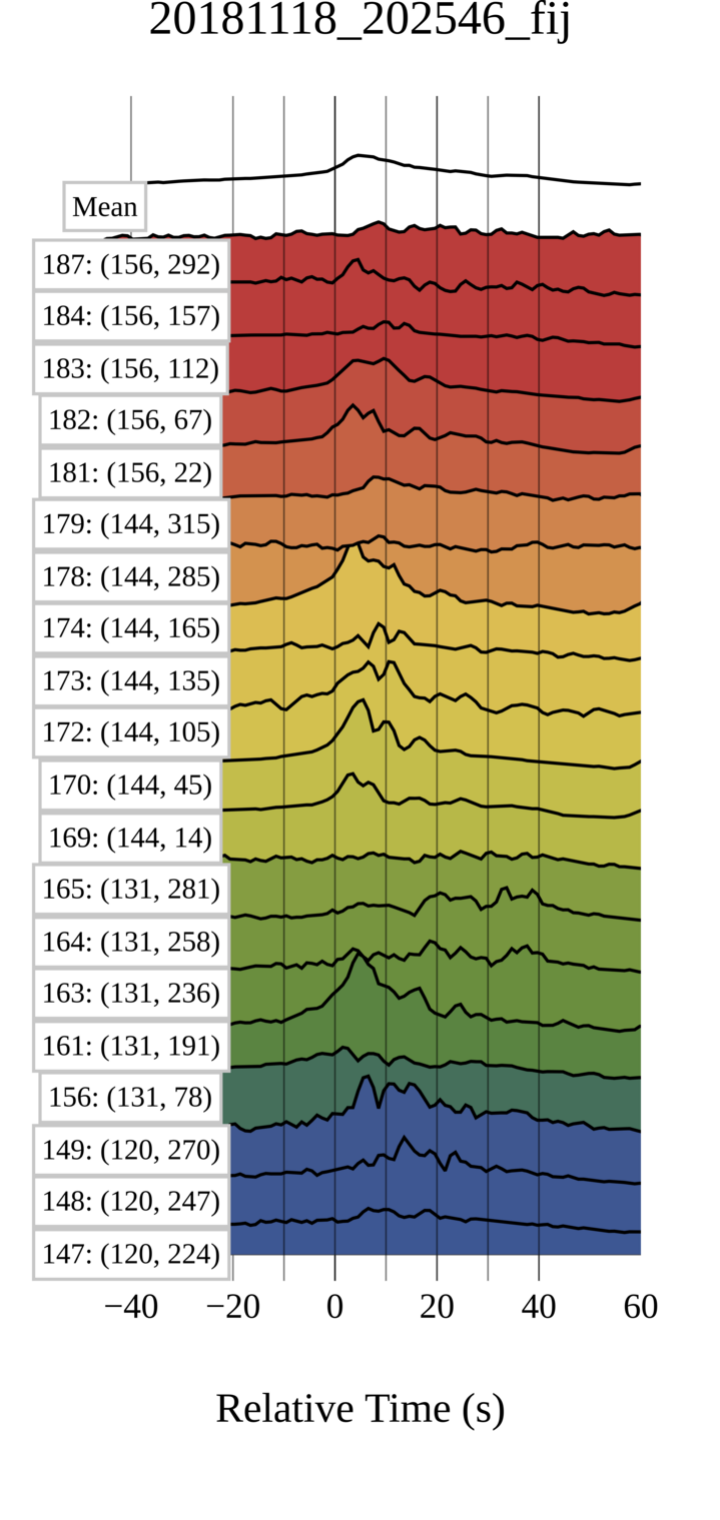}
         \caption{pP Envelope Stacking}
         \label{fig:2002_okt_rawpp}
     \end{subfigure}
     \begin{subfigure}[b]{0.225\textwidth}
         \centering
         \includegraphics[width=\textwidth]{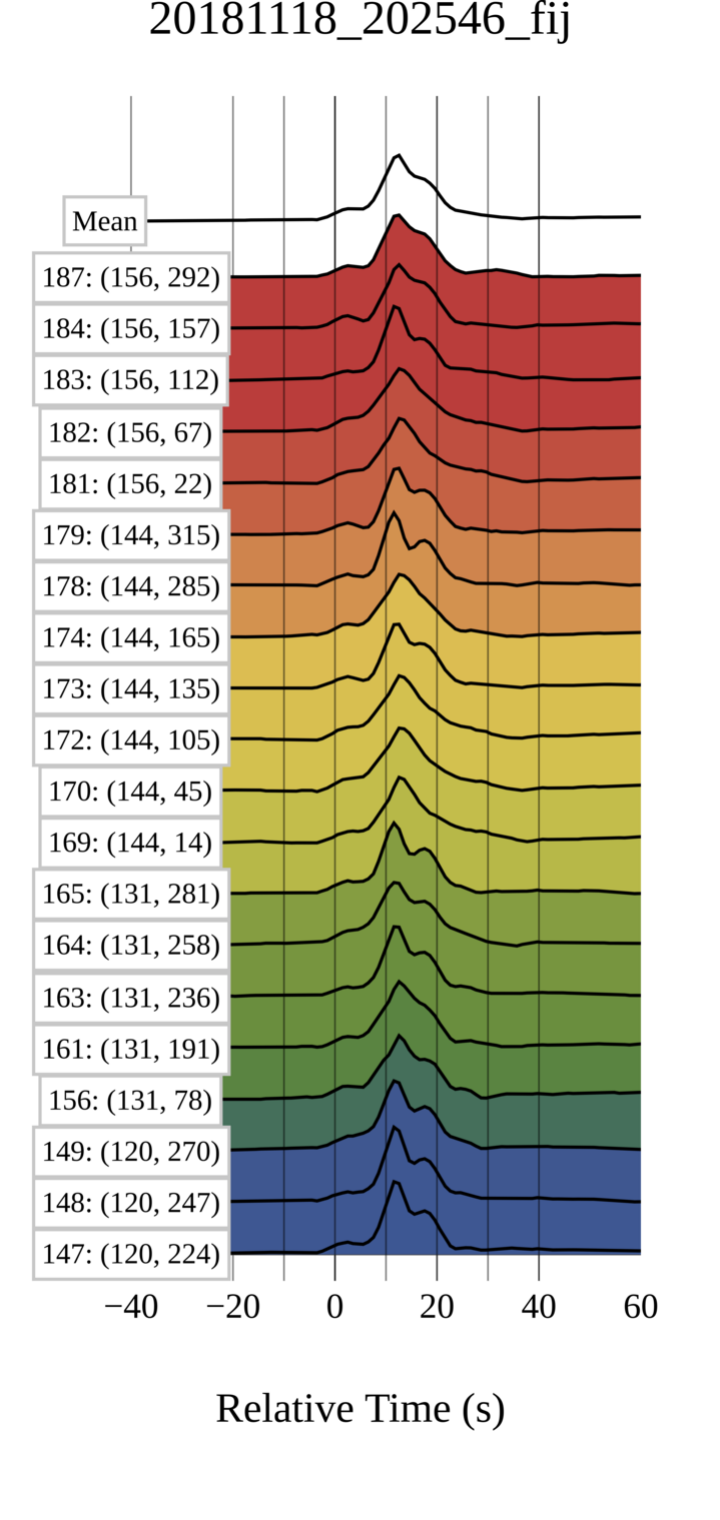}
         \caption{SymVAE pP STFs}
         \label{fig:2002_okt_usvspp}
     \end{subfigure}
        \caption{
        \label{fig:2018_fij}
        Analysis of P and pP directivity for the Mw 6.8 fij4 earthquake, where at least two closely spaced rupture episodes detectable in both P and pP STFs. 
        The resolution of pP STFs is reduced due to crustal attenuation. 
      Note the agreement between P and pP analyses, indicating that both the elongation and compression of the STFs are consistently detected across P and pP pixels.
       The fact that P and pP seismograms were independently trained using SymVAE for this earthquake offers strong validation for our approach. It is important to recognize that, due to phase uncertainties, the quality of pP envelope stacking is generally much lower than that of P.
        }
\end{figure}

SymVAE was separately trained using P-windowed and pP-windowed seismograms from all training earthquakes. The results of the fij4 earthquake near the Fiji Islands are presented in Fig.~\ref{fig:2018_fij}.
For deep seismic events, the raypaths of P and pP waves are predominantly independent, sampling different regions of the subsurface. The similarity in the characteristics of the STFs between the P and pP scenarios serves as a verification, confirming that the features identified by SymVAE are genuinely related to the earthquake source. Usually, the performance of the pP envelope stacking is poor compared to that of the P stacking, attributed to increased phase uncertainties arising from wave propagation through the crust.
For the fij4 earthquake, a significant similarity is identified between the P and pP STFs, despite the fact that the SymVAE models are independently trained for these phases. All STFs demonstrate two closely spaced episodes. Although the polar angles differ between the STFs of P and pP, one can approximately compare the P STFs with pP STFs that share a similar azimuthal angle to identify the correlation.
Note that pP STFs have lower resolution than P-wave STFs, suggesting that while SymVAE improves resolution better than envelope stacking, it cannot restore high-frequency details lost due to intrinsic attenuation --- this is also illustrated in our synthetic example. In this situation, pP experiences greater attenuation than P, as its energy traverses the crust twice, which inherently attenuates more.


\subsection{Consistency among pixels}
The principle of scale separation discussed in Section~\ref{sec:scale} implies that the source features should exhibit smooth variations between pixels. 
Furthermore, for earthquakes with smaller magnitudes and shorter fault lengths, the STFs are expected to be more homogeneous across pixels than in more complex and typically larger magnitude earthquakes.
Both expectations are met when analyzing the findings presented in this paper.
SymVAE independently estimates the STFs for each pixel. However, we observed
uniformity of the STFs among different pixels --- this is demonstrated for low magnitude earthquakes in Fig.~\ref{fig:bon2_mar}.
%
For earthquakes of higher magnitude in Fig.~\ref{fig:pb2_okt7}, there is more variation; however, there is a certain level of consistency between pixels. 
These observations suggest that the attributes captured by the coherent encoder of SymVAE are more strongly linked to the earthquake source than to any incoherent path influences. 
It is important to mention that, due to the non-uniform distribution of receivers, the pixel consistency was not observed with envelope stacks.

\section{Discussion}
Our approach relies on the scale separation between the source and path effects. This implies that there are limitations on the maximum fault dimensions and the range of wave frequencies that can be effectively analyzed. In cases of earthquakes with large fault dimensions, with frequencies higher than $0.1\,$Hz and near-field receiver locations, the Fraunhofer approximation is no longer valid. To address these scenarios, it is necessary to develop more advanced architectures for source imaging.
For shallow earthquakes, the interpretation of SymVAE-generated STFs must be done with care, especially considering that coherent scattering (e.g., surface-reflected phases) is not easily distinguishable from the source information~\citep{langston1978moments}.
We have demonstrated the applicability of SymVAE to both P and pP waveforms.
We propose a future generation of STFs for various time windows of far-field seismograms, offering the possibility of obtaining additional constraints on rupture parameters.
SymVAE generates STFs with superior resolution compared to conventional methods such as envelope stacking. 
These STFs have the potential to serve as valuable input for time-reversal methods, where they can simply be backpropagated in a homogeneous medium to image the source.


%
The issue with deep learning and generative models lies in the slight variations of generated source-time functions each time the network is trained. However, the interpretations presented in this paper remain consistent after multiple training sessions. Hyperparameter tuning is crucial for any deep learning model. Assessing the quality of generated seismograms is essential for fine-tuning network hyperparameters, and in this study, our tuning is guided by the sanity checks and synthetic tests outlined in the paper. 
Developing methods to verify the physical validity of virtual seismograms will significantly aid in the advancement of generative models.
Ultimately, the virtual seismograms created by SymVAE interpolate the measured seismograms in high dimensions. It is crucial to understand that we rely on this high-dimensional interpolation capability of networks, which benefits from training on a large number of earthquakes and is not necessarily guaranteed to be physically accurate. The architecture of both encoder and decoder networks affects the interpolation process. For example, when fully connected neural networks were used instead of convolutional networks, the quality of the generated STFs was unsatisfactory, implying that the inductive bias inherent in convolutional networks plays a significant role.
Similarly, we observed a significant drop in SymVAE's performance when trained on a single earthquake, highlighting the importance of using multiple earthquakes for training.


\section{Conclusions}

This paper addresses the characterization of deep-focus earthquakes using a novel method based on the variational symmetric autoencoder (SymVAE).
Unlike conventional methods that rely on a simple one-input, one-output convolutional model, SymVAE disentangles the source and path effects in a model-free fashion.
Our approach assumes scale separation,
where source processes operate on a slower spatial scale than path effects.
It is important to note that the SymVAE
training process is unsupervised and does not require labeled data, 
seismic arrival picking, or prior knowledge of subsurface scattering or source signature. Furthermore, the technique is highly scalable, enabling training on all available far-field seismograms from multiple earthquakes.
We showcased azimuth-specific band-limited source-time functions (STFs) derived from a series of deep earthquakes. We compared these derived STFs with raw displacement envelope stacking and traditional deconvolution methods.
The extracted STFs provide considerably enhanced resolution and emphasize weak secondary events.
Our results provide evidence that most deep earthquakes with Mw above 6.0 are characterized by fragmented seismic moment release with individual rutpure episodes of duration $5$--$10$s.
 %
%
In summary, our novel approach presents a promising direction for advancing the characterization of earthquake sources.

\section{Data Availability and Resources}
All data and codes used is this study are open access. 
Seismological data are sourced primarily from the Incorporated Research Institutions for Seismology Data Management Center (IRIS-DMC), where the data were acquired from a variety of institutions and seismic data networks,  \cite{doi_10_7914_SN_AE,doi_10_7914_SN_AI,doi_10_7914_SN_AK,doi_10_7914_SN_AT,doi_10_26186_144675,doi_10_7914_SN_AV,doi_10_14470_NJ617293,doi_10_7914_SN_AZ,doi_10_7914_SN_BE,doi_10_7932_BDSN,doi_10_7914_SN_BW,doi_10_7914_SN_BX,doi_10_7914_SN_C1,doi_10_7914_SN_CA,doi_10_7914_SN_CB,doi_10_7914_SN_CC,doi_10_12686_sed_networks_ch,doi_10_7914_SN_CI,doi_10_7914_SN_CM,doi_10_7914_SN_CN,doi_10_7914_SN_CR,doi_10_7914_SN_CU,doi_10_14470_PK615318,doi_10_7914_SN_CZ,doi_10_7914_SN_DR,doi_10_7914_SN_EI,doi_10_15778_RESIF_FR,doi_10_18715_GEOSCOPE_G,doi_10_7914_av8j_nc83,doi_10_14470_TR560404,doi_10_7914_SN_GG,doi_10_25928_mbx6_hr74,doi_10_7914_SN_GS,doi_10_7914_SN_GT,doi_10_7914_SN_GU,doi_10_14470_UR044600,doi_10_7914_SN_HL,doi_10_14470_UH028726,doi_10_7914_SN_IB,doi_10_7914_SN_IC,SN/AE,SN/AI,SN/AK,SN/AT,AU-ANSS,SN/AV,NJ617293,SN/AZ,SN/BE,BDSN,SN/BW,SN/BX,SN/C1,SN/CA,SN/CB,SN/CC,RESIF-CH}.
Each of these sources played an integral role in the assembly of the comprehensive dataset necessary to train symmetric autoencoders.
The training data and scripts
specifically used in the study will be made available using a github repository.
%
The half-duration values in Tab.~\ref{tab:eq_details} and the beachball plots in Fig.~\ref{fig:chlP} are taken from the GCMT (Global Centroid Moment Tensor) catalog.

\section{Acknowlegements}
This work is funded by the Science and Engineering Research Board, Department of Science and Technology, India (Grant Number SRG/2021/000205).
We thank Athira Vijayan help with data pre-processing. 
We also wish to express our appreciation for the insightful discussions held with Matthew Li and Laurent Demanet. 
We express our sincere gratitude to the anonymous reviewers for their thoughtful comments and constructive feedback from editors Martin Mai and Victor Tsai, which significantly improved the quality of this paper.
We extend our appreciation to the International Federation of Digital Seismograph Networks (FDSN) and IRIS Data Services for facilitating access to the earthquake waveform data.
Our research was significantly aided by the use of the Flux.jl deep learning framework, which enabled us to implement and train the SymVAE model. We also acknowledge the utilization of the Julia programming language, specifically version 1.11.0, as well as Pluto notebooks, which played a pivotal role in the development of our numerical simulations and data analysis presented throughout this work. We acknowledge the use of Plotly.js, a valuable tool that greatly facilitated the generation of the high-quality graphics and visualizations presented in this paper.
We also express our appreciation for PyGMT (GMT - Generic Mapping Tools).
Furthermore, we acknowledge the use of HealPix, a versatile and widely adopted pixelization scheme for mapping the spherical surface of the Earth.

\appendix
\appendixpage
\counterwithin{figure}{section}

\section{Fraunhofer Approximation}
\label{app:frf}
\begin{figure}
\noindent\includegraphics[width=\textwidth]{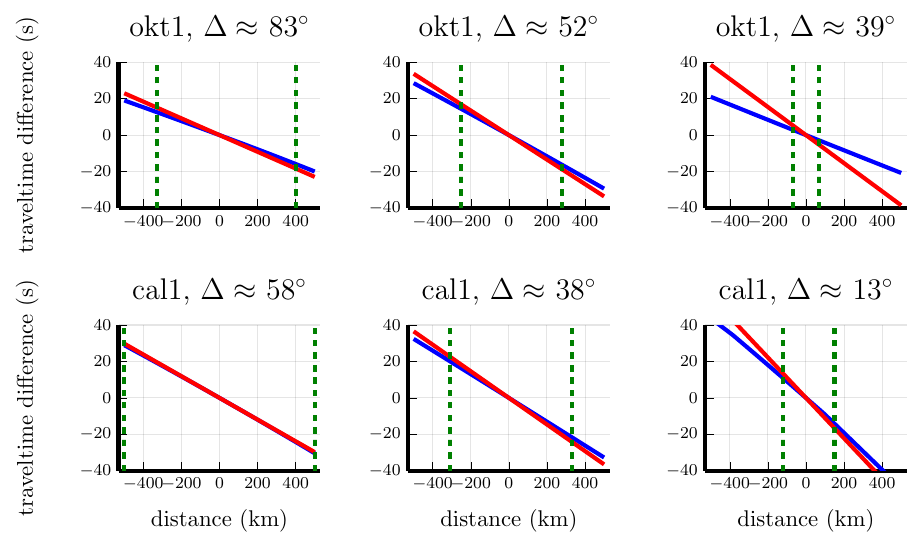}
\caption{
In this figure, red Fraunhofer-approximated traveltimes are compared to the actual IASP91 model 
traveltimes in blue, computed from different locations in the source region to a fixed receiver at a given epicentral distance. 
We considered a line source oriented in the direction of the ray leaving the focus. 
The approximation error increases as the epicentral distance decreases. The maximum fault length that can be analyzed based 
on the error in the Fraunhofer approximation is marked by dashed green lines, indicating the point where the error exceeds a quarter of the minimum period of 
10\,s considered in our analysis. In our work, we only used receivers with a minimum epicentral distance of 15$\degree$. 
Here, cal1 is a shallow crustal earthquake, while okt1 is a deep focus earthquake.
}
\label{fig:frf_src}
\end{figure}

We compare the travel times calculated using the Fraunhofer approximation with those obtained by tracing rays from various source locations to a receiver within the IASP91 model. Multiple receivers at epicentral distances greater than 10 degrees were considered.
The accuracy of the Fraunhofer approximation is contingent upon the introduced traveltime error, which should be no greater than one-fourth of the minimum period.
Consequently, the validity of this approximation is frequency-dependent. In our study, we analyzed a wavefield with limited bandwidth and restricted frequencies to a maximum of $0.1\,$Hz. This allowed the application of the Fraunhofer approximation for the lengths of dominant source region up to $300\,$ km, as shown in Figure \ref{fig:frf_src}. 
However, it should be noted that higher frequencies can still be analyzed when dealing with low-magnitude earthquakes that possess shorter dominant source region lengths.

\section{Generating Virtual Seismograms Using Source Effects from One Earthquake and Path Effects from Another}
\label{app:virtual}

In this appendix, we show that
SymVAE's latent representation of the seismograms can be manipulated to synthesize virtual or hybrid seismograms.
For instance, we can swap source and path codes of two different displacement seismograms, 
\begin{eqnarray}
\data{chl2}{k}{chl2}{A}\quad\text{and}\quad\data{chl3}{k}{chl3}{B}, 
\end{eqnarray}
to create two new virtual seismograms.
The first virtual seismogram,
\begin{eqnarray}
\label{eqn:app_virtual1}
\virtt{chl2}{k}{chl3}{B}&=&\f\left(\smean{chl2}{k}, \pmean{chl3}{k}{chl3}{B}, \phi\right),
\end{eqnarray}
is generated by combining the source code from the $k$th pixel of the chl2 earthquake with the path code from the second chl3 seismogram.
Similarly, the second virtual seismogram, 
\begin{eqnarray}
\label{eqn:app_virtual2}
\virtt{chl3}{k}{chl2}{A}&=&\f\left(\smean{chl3}{k}, \pmean{chl2}{k}{chl2}{A}, \phi\right),
\end{eqnarray}
is generated by combining the source code from the $k$th pixel of chl3 with the path code from the first chl2 seismogram. 
In this context, $\smean{chl2}{k}$ denotes the average value of the posterior distribution of the source $Q(\s \mid \Data{chl2}{k}, \theta_\s)$. Likewise, $\pmean{chl2}{k}{chl2}{A}$ and $\pmean{chl3}{k}{chl3}{B}$ represent the average values for the path posterior distributions $Q(\p \mid \data{chl2}{k}{chl2}{A}, \theta_{\p})$ and $Q(\p \mid \data{chl3}{k}{chl3}{B}, \theta_{\p})$, accordingly.
Again, recall that
in our notation, the subscript denotes the source information contained in the seismogram, whereas the superscript denotes the path that connects the source to the receiver.

Do virtual seismograms have physical significance? What methods can be used to assess their accuracy?
We follow a methodology closely aligned with the empirical Green's function (EGF) approach, focusing on shallow earthquakes originating from the same source region.
We examine three earthquakes: chl1 (2010-02-27; Mw 8.8), chl2 (2010-03-05; Mw 6.6), and chl3 (2010-03-16; Mw 6.7), all originating from a source region located near the central coast of Chile (see Tab.~\ref{tab:eq_details}).
The assumption of EGF suggests that the path effects are consistent between three seismograms, each corresponding to an earthquake recorded at the same station.
This implies that the accuracy of the virtual seismogram $\virtt{chl2}{k}{chl3}{A}$ can be evaluated by direct comparison to the observed seismogram $\data{chl2}{k}{chl2}{A}$.
Here, 
note that since the earthquakes are collocated, station A is associated with the same focal-sphere pixel, indexed $k$, for all three earthquakes.

\begin{figure}
    \begin{subfigure}{0.5\textwidth}
        \centering
        \includegraphics[width=\linewidth]{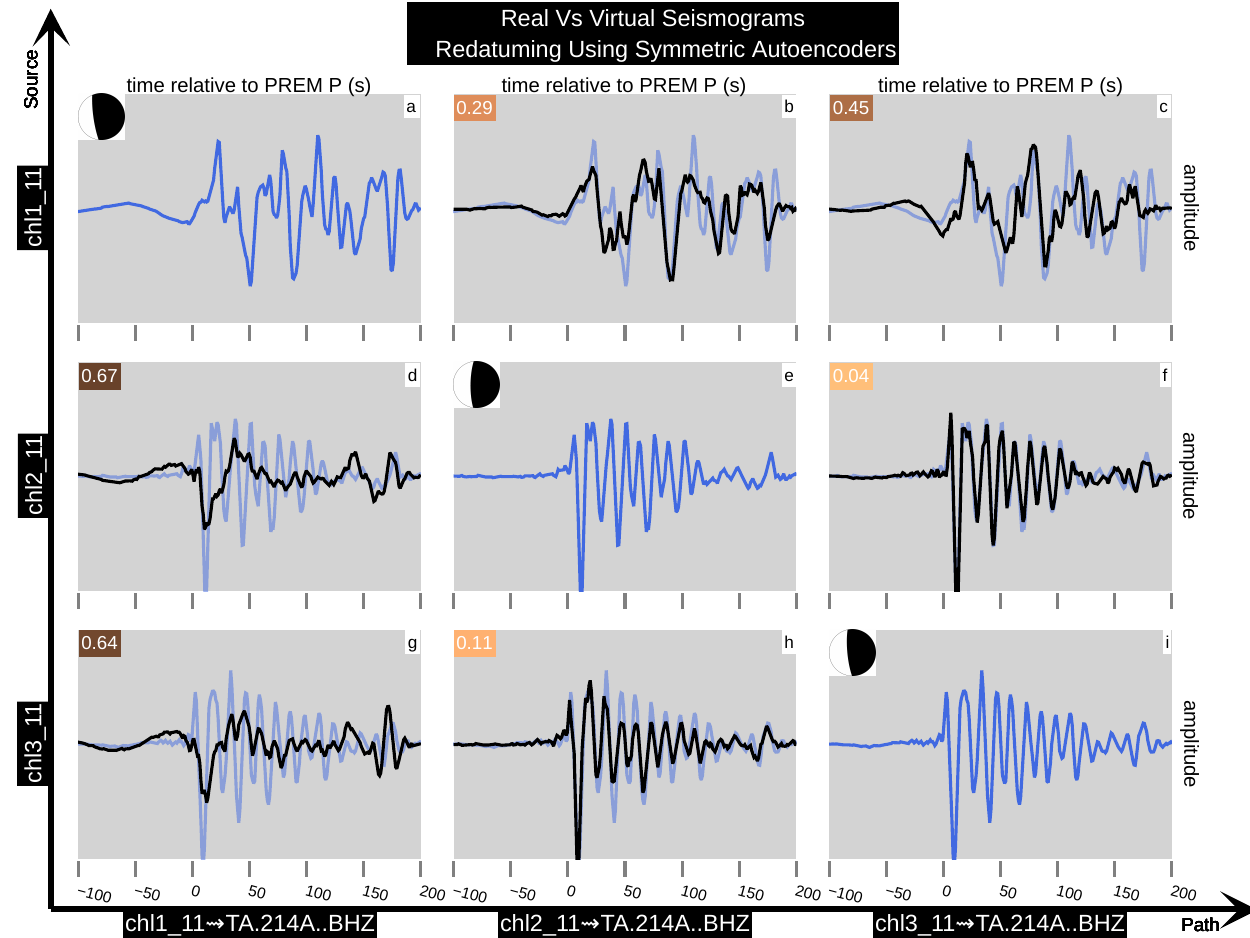}
        \caption{P Waveform Redatuming}
        \label{fig:chlP}
    \end{subfigure}
    \begin{subfigure}{0.5\textwidth}
        \centering
        \includegraphics[width=\linewidth]{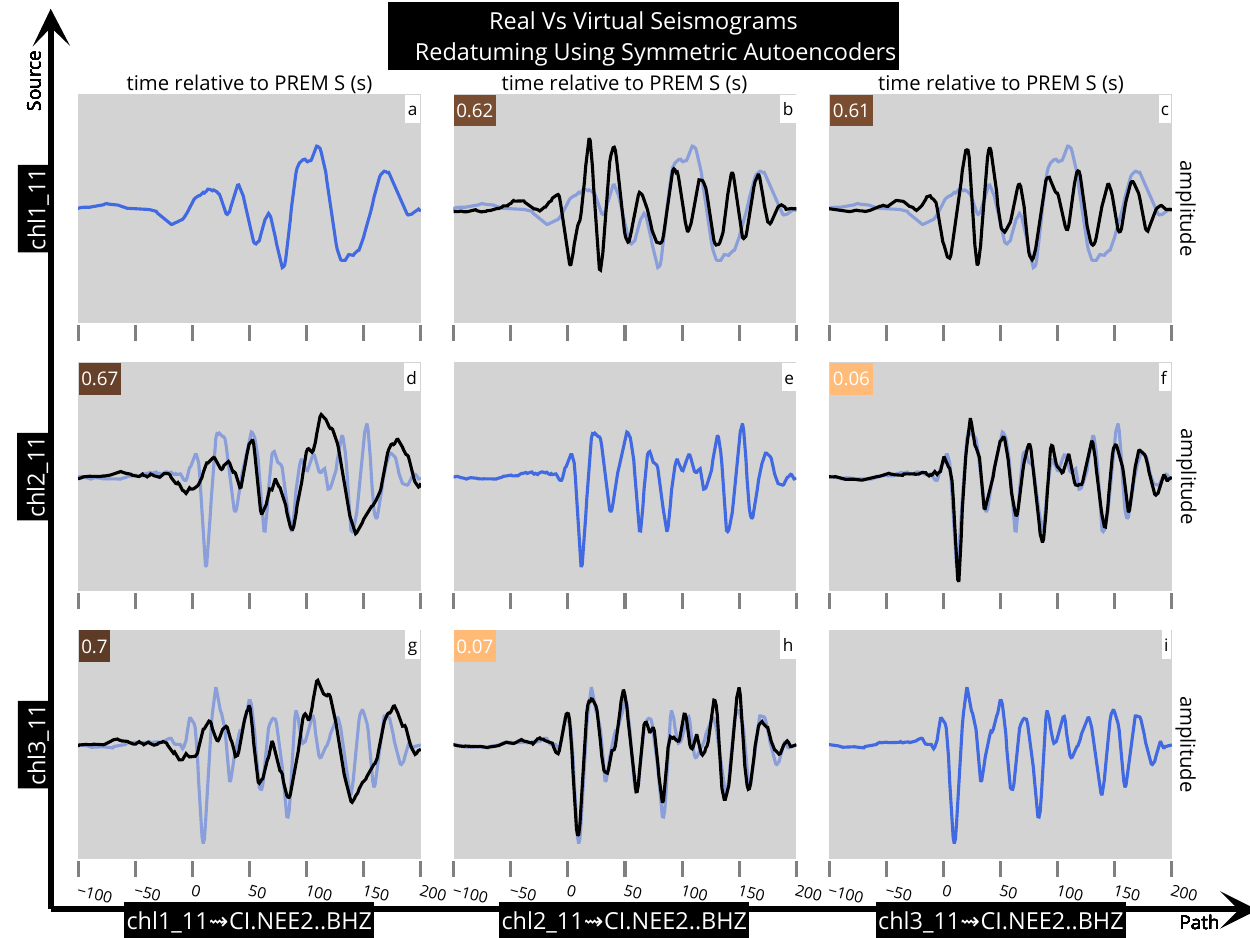}
        \caption{S Waveform Redatuming}
        \label{fig:chlS}
    \end{subfigure}
    \caption{
Displacement seismograms are embedded in the source-path latent space of SymVAE.
The blue (light shade) time series on the diagonal represent the measured seismograms, while the black (dark shade) ones are virtual seismograms generated through a process involving the exchange of source and path information between the measured seismograms. 
Since the earthquakes are selected from the same source region, if they share similarity, the virtual seismograms should exhibit strong correlation with the measured seismograms of a particular row.
In this context, the Pearson correlation distance between the measured and virtual seismograms is depicted.
(a) Demonstrates an example of P-windowed seismograms from earthquakes along the central coast of Chile. 
(b) Replicates the process in a), but with S-windowed seismograms.
In both scenarios, earthquakes chl2 and chl3 exhibit substantial similarity, whereas the presence of higher correlation distances implies a distinct source mechanism for chl1 when compared to chl2 and chl3.
There is an agreement between the P and S results. }
    \label{fig:app_chl}
\end{figure}

Next, we created six virtual seismograms by exchanging the path effects between each pair of the three earthquakes. 
All virtual seismograms are embedded in the latent space of SymVAE in Fig.~\ref{fig:app_chl} for both P and S arrivals. 
In this source path representation, the
seismograms plotted on the diagonal using blue color are observed, whereas the off-diagonal seismograms plotted in black are virtual seismograms.
To assess the error in a particular virtual seismogram, we calculated the Pearson correlation distance with the corresponding measured seismograms that share identical source information.
For example, virtual seismograms analyzing the lower-magnitude pair chl2 and chl3 show reduced error, since these earthquakes may possess simpler source mechanisms.
However, the
virtual seismograms generated using the path effects of the chl1 earthquake do not correlate with their respective measured seismograms. This discrepancy can be attributed to the fact that chl1 features more rupture episodes compared to the other two earthquakes.
We infer that it is feasible to create virtual seismograms by exchanging the source and path effects between similar earthquakes. The potential applications of these virtual seismograms for examining earthquake sources can be pursued in future studies.

\bibliographystyle{unsrtnat}
\bibliography{library, networks}  

\end{document}

%% file: notation.tex
\newcommand{\x}{\mathbf{x}}

\newcommand{\xs}{\boldsymbol{\xi}}

\newcommand{\source}[2]{\text{#1}\_{#2}}
\newcommand{\srpath}[2]{\text{#1}\rightsquigarrow #2}
\newcommand{\data}[4]{\mathbf{x}^{\srpath{#3}{#4}}_{\source{#1}{#2}}}
\newcommand{\Data}[2]{\mathbf{X}_{\source{#1}{#2}}}
\newcommand{\stf}[2]{\mathbf{y}_{\source{#1}{#2}}}
\newcommand{\Stf}[1]{\mathbf{Y}_{\text{#1}}}
\newcommand{\virt}[3]{\mathbf{y}^{#3}_{\source{#1}{#2}}}
\newcommand{\virtt}[4]{\mathbf{y}^{\srpath{#3}{#4}}_{\source{#1}{#2}}}

\newcommand{\smean}[2]{\mathbf{s}_{\source{#1}{#2}}}
\newcommand{\pmean}[4]{\mathbf{p}^{\srpath{#3}{#4}}_{\source{#1}{#2}}}



%% file: tablep2.tex
\begin{tabular}{ccc}
\makecell{dominant \\fault length\\ (km)} & \makecell{angular \\ resolution \\ (degree)}  & \makecell{angular \\ resolution \\ (degree)} \\
$L_{\text{max}}$ & $\Delta\psi$ for P waves & $\Delta\psi$ for S waves\\
100.00 & 103.1 & 51.6\\
200.00 & 51.5 & 25.8\\
300.00 & 34.4 & 17.2\\
400.00 & 25.8 & 13.1\\
\end{tabular}

%% file: new_table.tex
\begin{tabular}{|cccccc|}
\toprule
\hline
\# & Code &       EQ Name &  Mw &  Depth\,(km) & Region \\ \hline
\midrule
     1 & nbl & 19940609\_003316\_nbl & 8.20 & 631.3 &              Northern Bolivia \\
     2 & okt7 & 20021117\_045353\_okt & 7.3 &  459.1 &    Sea of Okhotsk \\
     3 & okt3 & 20080705\_021204\_okt & 7.7 &   632.8 &   Sea of Okhotsk \\
    4 & okt4 & 20081124\_090258\_okt & 7.3 & 492 & Sea of Okhotsk\\
    5 &  spn & 20100411\_220812\_spn & 6.30 & 609.8 &                     Spain \\
    6 & okt6 & 20120814\_025938\_okt & 7.70 & 583.00 &          Sea of Okhotsk \\
    7 & okt1 & 20130524\_054448\_okt & 8.3 & 598 &           Sea of Okhotsk \\
    8 & okt2 & 20130524\_145631\_okt & 6.7 & 624 & Sea of Okhotsk\\
    9 & okt5 & 20131001\_033821\_okt & 6.7 & 573 & Sea of Okhotsk\\
    10 & van1 & 20140305\_095657\_van & 6.30 & 638.00 &           Vanuatu Islands \\
    11 & fij3 & 20141101\_185722\_fij & 7.10 & 434.00 &               Fiji Islands  \\
    12 & bon1 & 20150530\_112302\_bon & 7.80 & 664.00 &      Bonin Island Japan Reg \\
    13 &  pb2 & 20151124\_224538\_pb2 & 7.60 & 606.00 &                Peru Brazil \\
    14 & fij1 & 20180819\_001940\_fij & 8.20 & 600.00 &             Fiji Islands  \\
    15 & fij2 & 20180906\_154918\_fij & 7.90 & 670.00 &             Fiji Islands  \\
    16 & fij4 & 20181118\_202546\_fij & 6.80 & 540.00 &              Fiji Islands  \\
    17 & sol & 20190126\_035137\_sol & 6.2 & 355 & Solomon Islands\\
    18 & bnd & 20190406\_215501\_bnd & 6.3 & 539 & Banda Sea\\
    19 & hns & 20190604\_043916\_hns & 6.3 & 430.3 & Southeast of Honshu, Japan\\
    20 & mar & 20190628\_155131\_mar & 6.4 & 410 & Mariana Islands\\
    21 & sol & 20190711\_170837\_sol & 6.0 & 495.17 & Solomon Islands\\
    22 & fij5 & 20210424\_002338\_fij & 6.50 & 301.00 &            Fiji Islands  \\
    23 & jav & 20230414\_095545\_jav & 7.0 & 594 & Java, Indonesia\\
    24 & san & 20230805\_072006\_santiago & 6.2 & 575 & Santiago Del Estero Province, Argentina\\
    25 & bal & 20230828\_195530\_bal & 7.1 & 500 & Bali Sea\\
    26 & fij & 20231031\_111056\_fij & 6.5 & 560 & Fiji Islands Region\\
    27 & brz1 & 20240120\_213105\_brz & 6.6 & 607 & Western Brazil\\
    28 & brz2 & 20240128\_093856\_brz & 6.5 & 621 & Western Brazil\\
    29 & bon2 & 20240427\_083534\_bon & 6.5 & 496  & Bonin Islands, Japan\\
    30 & bon3 & 20240707\_200112\_bon & 6.2 & 546 & Bonin Islands, Japan\\
    31 & php & 20240711\_021318\_php & 7.1 & 640 & Mindanao, Philippines\\
    32 & okt8 & 20240810\_032832\_okt & 6.5 & 407 & Sea of Okhotsk\\
    \hline
\bottomrule
\end{tabular}